\documentclass[a4paper,11pt]{article}

\usepackage{hyperref}%
\hypersetup{%
  colorlinks=true,%
  linkcolor=blue,%
  citecolor=magenta,%
  urlcolor=black,%
  bookmarksnumbered=true,%
  bookmarkstype=toc,%
  bookmarksopen=false,%
  pdftitle={ERG Kernels on Multiply Connected Configuration Spaces},%
  pdfkeywords={exact renormalization group; topology; Wilson loop},%
  pdfauthor={Satoshi Ohya and Taichi Tanaka}}%

\usepackage[margin=1in]{geometry}%

\usepackage[tt=false,sb]{libertine}%
\usepackage[T1]{fontenc}%
\usepackage{libertinust1math}%
\usepackage[cal=stix,scr=boondoxo,bb=boondox]{mathalpha}%
\usepackage{bm}%
\usepackage{mathtools}%
\DeclareMathOperator{\Map}{Map}%
\DeclareMathOperator{\e}{e}%
\DeclareMathOperator{\id}{id}%
\DeclareMathOperator{\vol}{vol}%
\DeclareMathOperator{\Isom}{Isom}%
\DeclareMathOperator{\tr}{tr}%
\allowdisplaybreaks[1]%

\usepackage{enumitem}

\usepackage{tikz}%

\usepackage[font=small,hang]{caption}%
\usepackage[labelformat=simple]{subcaption}%
\usepackage{amsthm}%
\usepackage[many]{tcolorbox}%
\newtheorem*{theorem*}{Theorem}%
\tcolorboxenvironment{theorem*}{%
  colback=black!10!white,%
  colframe=black!10!white,%
  boxrule=0pt,%
  boxsep=0pt,%
  left=1pt,%
  right=1pt,%
  top=1pt,%
  bottom=1pt,%
  oversize=0pt,%
  sharp corners,%
}%

\usepackage[title]{appendix}%

\usepackage[backend=biber,style=phys,biblabel=brackets,block=ragged,eprint=true,url=true]{biblatex}%
\title{\Large\bfseries ERG Kernels on Multiply Connected Configuration Spaces}%
\author{\normalsize Satoshi Ohya${}^{1,}$\footnote{\texttt{ohya.satoshi@nihon-u.ac.jp}}~~~and~~~Taichi Tanaka${}^{2,}$\footnote{\texttt{csti24001@g.nihon-u.ac.jp}}\\[1em]
  \small\itshape ${}^{1}$Institute of Quantum Science, Nihon University,\\
  \small\itshape Kanda-Surugadai 1-8-14, Chiyoda, Tokyo 101-8308, Japan\\
  \small\itshape ${}^{2}$Graduate School of Science and Technology, Nihon University,\\
  \small\itshape Kanda-Surugadai 1-8-14, Chiyoda, Tokyo 101-8308,
  Japan}%
\date{\small(Dated: \today)}%

\begin{document}
\maketitle%
\flushbottom%

\begin{abstract}
  In the functional-integral formulation of Euclidean field theory,
  exact renormalization group (ERG) transformations are realized by
  functional-integral kernels. Unlike the ERG flow equations that
  describe infinitesimal ERG transformations, these ERG kernels
  explicitly depend on the global topology of the configuration
  space. This paper explores this topology dependence for multiply
  connected configuration spaces. We show that the ERG kernel is in
  general given by a weighted sum of kernels on its universal covering
  space, where the weight factors are determined by a one-dimensional
  representation of the fundamental group. These weight factors are
  shown never to be renormalized under the ERG. We also show that
  these factors can be interpreted as Aharonov-Bohm phases with
  respect to a background magnetic flux penetrating the
  infinite-dimensional configuration space. From this viewpoint, a
  normalization condition for ERG transformations corresponds to a
  flux-quantization condition, which is equivalent to the
  level-quantization condition for Wess-Zumino-Witten terms in
  nonlinear sigma models. Finally, we present an alternative
  gauge-equivalent form of the ERG flow equation that incorporates
  this topological information locally.
\end{abstract}

\newpage
\begingroup%
\hypersetup{linkcolor=black}%
\setcounter{tocdepth}{2}
\tableofcontents%
\endgroup%

\section{Introduction}
\label{section:1}
The topology of the configuration space plays an important role in
field theory. In general, a field $\phi$ in the functional-integral
formulation of Euclidean field theory is a map from a spacetime $X$ to
a target space $Y$ with or without boundary conditions. Consider, for
example, Euclidean field theory with the Dirichlet boundary condition
$\phi(x_{\ast})=y_{\ast}$, where $x_{\ast}\in X$ and $y_{\ast}\in Y$
are specifically chosen points called \textit{base points}. In this
case, the configuration space $Q$ of Euclidean field theory can be
regarded as the following \textit{based mapping space}:
\begin{align}
  Q
  =\Map_{\ast}(X,Y)
  =\{\phi:X\to Y\mid\phi(x_{\ast})=y_{\ast}\}.\label{eq:1}
\end{align}
Without particular boundary conditions, on the other hand, the
configuration space $Q$ can simply be regarded as the following
(unbased) \textit{mapping space}:
\begin{align}
  Q
  =\Map(X,Y)
  =\{\phi:X\to Y\}.\label{eq:2}
\end{align}

As is well known in algebraic topology, the (un)based mapping spaces
\eqref{eq:1} and \eqref{eq:2} can be topologically nontrivial. For
instance, consider the $O(n)$ nonlinear sigma model on the
$d$-dimensional Euclidean spacetime $\mathbb{R}^{d}$, where the target
space is the $(n-1)$-sphere $S^{n-1}$. Suppose that the field $\phi$
fulfills the isotropic boundary condition at infinity,
$\lim_{|x|\to\infty}\phi(x)=y_{\ast}$, where $y_{\ast}\in S^{n-1}$ is
a specific point in the target space. Under this boundary condition,
the domain of $\phi$ effectively contains the point at infinity such
that the spacetime effectively becomes the one-point compactification
of $\mathbb{R}^{d}$, which is topologically equivalent to the
$d$-sphere $S^{d}$. Thus we have
$X=\mathbb{R}^{d}\cup\{\infty\}\approx S^{d}$ and $Y=S^{n-1}$, where
$x_{\ast}=\infty\in X$ and $y_{\ast}\in Y$ play the roles of base
points. In this case, the configuration space of the $d$-dimensional
$O(n)$ nonlinear sigma model can be regarded as the based mapping
space $Q=\Map_{\ast}(S^{d},S^{n-1})$, whose $k$th homotopy group is
given by the $(k+d)$th homotopy group of $S^{n-1}$:\footnote{This just
  follows from the homeomorphism
  $\Map_{\ast}(X,\Map_{\ast}(Y,Z))\approx\Map_{\ast}(X\wedge Y,Z)$,
  where $\wedge$ stands for the smash product. By choosing $X=S^{k}$,
  $Y=S^{d}$, $Z=S^{n-1}$ and using the homeomorphism
  $S^{k}\wedge S^{d}\approx S^{k+d}$, we find
  $\Map_{\ast}(S^{k},\Map_{\ast}(S^{d},S^{n-1}))\approx\Map_{\ast}(S^{k+d},S^{n-1})$.
  This proves the isomorphism \eqref{eq:3} because the $k$th homotopy
  group $\pi_{k}(\Map_{\ast}(S^{d},S^{n-1}))$ is just the quotient
  space $\Map_{\ast}(S^{k},\Map_{\ast}(S^{d},S^{n-1}))/\!\sim$ under
  the base-point-preserving homotopy. See also Appendix of
  \cite{Kobayashi:1997rf}.}
\begin{align}
  \pi_{k}(Q)\cong\pi_{k+d}(S^{n-1}).\label{eq:3}
\end{align}
In particular, we have $\pi_{1}(Q)\cong\pi_{d+1}(S^{n-1})$. It then
follows from the homotopy groups of spheres (see, e.g., Section 4.1 of
\cite{Hatcher:2001}) that the configuration space $Q$ of the $O(n)$
nonlinear sigma model becomes multiply connected if
$(d,n)=(1,3),(2,3),(2,4),(3,3),(3,4),(3,5)$, and so on. Notice that if
$\pi_{1}(Q)$ is nontrivial, path integrals on $Q$ should be performed
on each homotopically distinct sector separately, because paths cannot
be continuously deformed from one sector to another.\footnote{This
  assumes that $\pi_{0}(Q)$ is trivial. If $\pi_{0}(Q)$ is nontrivial,
  $Q$ is not path-connected and can be decomposed into a disjoint
  union of path-connected components. In such cases, functional
  integrals must be evaluated on each path-connected component
  separately. However, as long as we focus on a single path-connected
  component of $Q$, our formalism remains applicable even when both
  $\pi_{0}(Q)$ and $\pi_{1}(Q)$ are nontrivial. A typical example of
  this is the two-dimensional $O(3)$ nonlinear sigma model, where
  $\pi_{0}(Q)\cong\pi_{2}(S^{2})\cong\mathbb{Z}$ and
  $\pi_{1}(Q)\cong\pi_{3}(S^{2})\cong\mathbb{Z}$. For simplicity, we
  assume throughout this paper that $\pi_{0}(Q)$ is trivial. Note that
  a nontrivial $\pi_{0}(Q)$ indicates the existence of instanton
  sectors.}

In general, functional integrals with nontrivial $\pi_{1}(Q)$ are best
described by Dowker's \textit{covering-space method}
\cite{Dowker:1972np}. In this method, we first perform the functional
integral on the universal covering space $\widetilde{Q}$ of $Q$, and
then make identifications under the deck transformation group
$\Gamma\cong\pi_{1}(Q)$, and then sum up all the identified
contributions.\footnote{The deck transformation group $\Gamma$ is a
  discrete subgroup of the homeomorphism group of $\widetilde{Q}$. It
  acts on $\widetilde{Q}$ with no fixed point and is isomorphic to the
  fundamental group $\pi_{1}(Q)$. Once we know $\widetilde{Q}$ and
  $\Gamma$, we can reconstruct $Q$ as the orbit space
  $Q\approx\widetilde{Q}/\Gamma$. For more details on these
  mathematical notions, we refer to Section 1.3 of Hatcher's textbook
  \cite{Hatcher:2001}.}  Note that this method is quite versatile and
applicable to any integral transforms on
$Q\approx\widetilde{Q}/\Gamma$. Hence, the covering-space method can
be applied to the \textit{exact renormalization group} (ERG) as well,
because the ERG transformation \textit{\`{a} la} Polchinski
\cite{Polchinski:1983gv} is a functional-integral transform for
Boltzmann weights. To the best of our knowledge, however, the topology
dependence of finite ERG transformations and its physical consequences
have not been elucidated before.

The purpose of this paper is to develop the ERG transformation in
Euclidean field theory whose configuration space $Q$ is multiply
connected. As we will show below, the functional-integral kernel of
the ERG transformation---the \textit{ERG kernel}---is in general given
by a weighted sum of the ERG kernels on the universal covering space
$\widetilde{Q}$, where weight factors are given by a one-dimensional
representation of $\Gamma\cong\pi_{1}(Q)$. By using this result, we
present a simple nonperturbative proof for the nonrenormalization of
Wess-Zumino-Witten terms in nonlinear sigma models.

The rest of the paper is organized as follows. In section
\ref{section:2}, we first review the basics of the covering-space
method by using quantum mechanics on a circle. We then present the
covering-space construction of the ERG kernel in section
\ref{section:3}. Section \ref{section:4} discusses the Wilson-loop
realization of the weight factors as well as alternative
gauge-equivalent forms of the ERG kernel and the flow equation. We
conclude in section \ref{section:5}. Appendix \ref{appendix:A}
presents some explicit examples of the ERG kernels.

\section{A prototypical example of the covering-space method: Quantum
  mechanics on
  \texorpdfstring{$S^{1}\approx\mathbb{R}/\mathbb{Z}$}{S1=R/Z}}
\label{section:2}
Dowker's covering-space method \cite{Dowker:1972np} is best understood
through the example of one-particle quantum mechanics on a circle. In
this section, we will illustrate the basic ideas of the method by
using the time-evolution kernel.

Let us consider quantum mechanics for a single spinless particle on a
circle $S^{1}$ of circumference $L$. Let $\psi(x)$ be a wavefunction
of the system, where $0\leq x< L$. We wish to find the time evolution
of the state $\psi$ by an amount of time $t$, which is given by the
unitary map
\begin{align}
  U_{t}:\psi\mapsto\psi_{t},\label{eq:4}
\end{align}
where $\psi_{t}$ is defined by the following integral transform:
\begin{align}
  \psi_{t}(x)=\int_{0}^{L}\!\!dy\,U_{t}(x,y)\psi(y).\label{eq:5}
\end{align}
Here $U_{t}(\ast,\ast)$ is the time-evolution kernel---the integral
kernel of the integral transform \eqref{eq:5}.\footnote{In Dirac's
  bracket notation,
  $U_{t}(x,y)=\langle x|\e^{-\frac{i}{\hbar}Ht}|y\rangle$, where $H$
  is the Hamiltonian of the system.} Once we know this kernel, we can
construct the time-evolved state $\psi_{t}$ for any initial state
$\psi$. Below we will construct the general form of the time-evolution
kernel on $S^{1}$ by using the covering-space method. This method
consists of the following three steps.

The first step is to forget about the circle $S^{1}$ for a moment;
instead, we consider the whole real line $\mathbb{R}$. Note that the
relation between $\mathbb{R}$ and $S^{1}$ is the quotient
$S^{1}\approx\mathbb{R}/\mathbb{Z}$, where $\mathbb{Z}$ is the
additive group of integers whose action on $\mathbb{R}$ is defined by
the discrete translation $x\mapsto x+nL$ ($n=0,\pm1,\cdots$). In fact,
by introducing the equivalence relation $\sim$ on $\mathbb{R}$ by
$x\sim x+nL$, we find $S^{1}$ is equivalent to the quotient space
(orbit space) $\mathbb{R}/\mathbb{Z}$. In the homotopy language,
$\mathbb{R}$ is the universal covering space of $S^{1}$ and the
discrete translation is the deck transformation on the universal
covering space $\mathbb{R}$, which forms the discrete group
$\mathbb{Z}$ called the deck transformation group that is isomorphic
to the fundamental group $\pi_{1}(S^{1})$. In short, the first step of
the covering-space method is to identify the universal covering space
and the deck transformation group.

The second step is to introduce an equivalence relation in the space
of wavefunctions on $\mathbb{R}$. We wish to physically identify a
wavefunction $\psi(x)$ at $x$ with a wavefunction $\psi(x+nL)$ at
$x+nL$. Note that $\psi(x)$ and $\psi(x+nL)$ are said to be physically
equivalent if their squared moduli (i.e., local probability densities)
are the same. So we require
\begin{align}
  |\psi(x+nL)|^{2}=|\psi(x)|^{2}\label{eq:6}
\end{align}
for arbitrary $x\in\mathbb{R}$ and $n\in\mathbb{Z}$. The solution to
the condition \eqref{eq:6} is
\begin{align}
  \psi(x+nL)=D(n)\psi(x),\label{eq:7}
\end{align}
where $D(n)\in U(1)$ is a pure phase that may depend on $n$. Note that
this phase cannot be arbitrary. To see this, consider the following
trivial identity for any $n,m\in\mathbb{Z}$:
\begin{align}
  \psi(x+(n+m)L)=\psi((x+mL)+nL).\label{eq:8}
\end{align}
It follows from \eqref{eq:7} that the left-hand side is evaluated as
$\psi(x+(n+m)L)=D(n+m)\psi(x)$. On the other hand, the right-hand side
gives $\psi((x+mL)+nL)=D(n)\psi(x+mL)=D(n)D(m)\psi(x)$. Hence we must
have $D(n+m)\psi(x)=D(n)D(m)\psi(x)$ for any $\psi$. Thus we find
\begin{align}
  D(n+m)=D(n)D(m).\label{eq:9}
\end{align}
This equation says that $D$ must be a map $D:\mathbb{Z}\to U(1)$ that
preserves the group multiplication rule; that is, $D$ must be a
one-dimensional unitary representation (or character) of
$\mathbb{Z}$. As is well known, there is a one-parameter family of
such unitary representations labeled by an angle parameter. The result
is
\begin{align}
  D(n)=\e^{in\theta},\label{eq:10}
\end{align}
where $\theta\in[0,2\pi)$ is a representation label. In short, the
second step of the covering-space method is to identify the
equivalence relation for wavefunctions that is compatible with the
deck transformation group. In quantum mechanics, the result is the
twisted boundary condition \eqref{eq:7} described by the
one-dimensional unitary representation \eqref{eq:10}. The origin of
this is just the phase ambiguity of wavefunctions.

Finally, the third step of the covering-space method is to consider
the time-evolution on the universal covering space $\mathbb{R}$ rather
than $S^{1}$. Let $\widetilde{U}_{t}(\ast,\ast)$ be a time-evolution
kernel on $\mathbb{R}$ satisfying
\begin{align}
  \psi_{t}(x)=\int_{-\infty}^{\infty}\!dy\,\widetilde{U}_{t}(x,y)\psi(y).\label{eq:11}
\end{align}
Once we know such a kernel $\widetilde{U}_{t}(\ast,\ast)$, we can
construct the time-evolution kernel on $S^{1}$. The key is the
following identity:
\begin{align}
  \int_{-\infty}^{\infty}\!\!dx\,f(x)=\int_{0}^{L}\!\!dx\sum_{n=-\infty}^{\infty}f(x+nL),\label{eq:12}
\end{align}
where $f$ is an arbitrary (integrable) function. This identity just
says that first summing over the orbit
$\mathbb{Z}x=\{x+nL\mid n\in\mathbb{Z}\}$ of $x\in[0,L)$ and then
integrating over the fundamental domain
$\mathbb{R}/\mathbb{Z}=\{x\mid0\leq x< L\}$ of $x$ gives the
integration over the whole universal covering space $\mathbb{R}$. It
is now easy to see that \eqref{eq:11} can be rewritten as follows:
\begin{align}
  \psi_{t}(x)
  &=\int_{-\infty}^{\infty}\!dy\,\widetilde{U}_{t}(x,y)\psi(y)\nonumber\\
  &=\int_{0}^{L}\!dy\sum_{n=-\infty}^{\infty}\widetilde{U}_{t}(x,y+nL)\psi(y+nL)\nonumber\\
  &=\int_{0}^{L}\!dy\sum_{n=-\infty}^{\infty}\widetilde{U}_{t}(x,y+nL)D(n)\psi(y)\nonumber\\
  &=\int_{0}^{L}\!dy\left(\sum_{n=-\infty}^{\infty}D(n)\widetilde{U}_{t}(x,y+nL)\right)\psi(y),\label{eq:13}
\end{align}
where the second equality follows from the identity \eqref{eq:12} and
the third equality follows from the twisted boundary condition
\eqref{eq:7}. Comparing the last line of \eqref{eq:13} with
\eqref{eq:5}, we finally obtain
\begin{align}
  U_{t}(x,y)=\sum_{n=-\infty}^{\infty}D(n)\widetilde{U}_{t}(x,y+nL).\label{eq:14}
\end{align}
This is the general form of the time-evolution kernel for one-particle
quantum mechanics on $S^{1}$. Note that $U_{t}(\ast,\ast)$ and
$\widetilde{U}_{t}(\ast,\ast)$ satisfy the same Schr\"{o}dinger
equation because \eqref{eq:14} is just the linear combination. Hence,
by solving the Schr\"{o}dinger equation
$i\hbar\frac{\partial}{\partial
  t}\widetilde{U}_{t}(x,y)=H\widetilde{U}_{t}(x,y)$ under the initial
condition $\widetilde{U}_{0}(x,y)=\delta(x-y)$ on $\mathbb{R}$, one
can construct the time-evolution kernel on $S^{1}$ by \eqref{eq:14}.

To recapitulate, an integral kernel on a multiply connected space is
given by the integral kernel on its universal covering space that
satisfies the same linear differential equation as the original
one. The final result is a linear combination like \eqref{eq:14},
where the linear-combination coefficients---the \textit{weight
  factors}---are determined by representation theory of the deck
transformation group. This is Dowker's covering-space method and can
be applied to the ERG as well. There, the time-evolution kernel
corresponds to the ERG kernel, the Schr\"{o}dinger equation
corresponds to the flow equation, the equivalence of local probability
densities corresponds to the equivalence of canonical distributions,
and the phase ambiguity of wavefunctions corresponds to the
constant-term ambiguity of Euclidean actions. Let us next go through
these things by moving on to Euclidean field theory on multiply
connected configuration space.

\section{ERG kernels on multiply connected configuration spaces}
\label{section:3}
Loosely speaking, Euclidean field theory is an infinite-dimensional
version of classical statistical mechanics, or classical probability
theory, where a random variable is a field $\phi$, the probability
distribution is the canonical distribution\footnote{The canonical
  distribution can be complex in Euclidean field theory. So it is not
  literally a probability distribution. However, to fulfill the
  reflection positivity, the canonical distribution must be reflection
  real; that is, it must be invariant under the composite operation of
  complex conjugation and orientation reversal. See section
  \ref{section:3.3}.}, and the moments with respect to the canonical
distribution are correlation functions; see, e.g., Section 1.5 of
\cite{Fernandez:1992jh}. If the correlation functions satisfy the
Osterwalder-Schrader axioms
\cite{Osterwalder:1973dx,Osterwalder:1974tc}, they can be analytically
continued to Wightman functions that satisfy the Wightman axioms of
Lorentzian quantum field theory. Once we obtain such Wightman
functions, we can reconstruct a Hilbert space, a vacuum state, and
field operators via the Wightman reconstruction theorem
\cite{Streater:1989vi}. This is the general story for obtaining
Lorentzian quantum field theory from Euclidean field theory
\cite{Glimm:1987ylb}, and the ultimate goal of the ERG is to construct
correlation functions satisfying the Osterwalder-Schrader axioms by
taking the continuum limit of correlation functions in cutoff field
theory. In Polchinski's formulation of the ERG
\cite{Polchinski:1983gv}, however, this continuum limit is formulated
without direct recourse to correlation functions.

Let $S_{\text{bare}}[\phi]$ be a bare action with respect to which we
wish to establish the continuum limit. This action depends on the
ultraviolet momentum cutoff $\Lambda_{0}(>0)$ as well as a set of bare
coupling constants. Instead of studying the correlation functions with
respect to $S_{\text{bare}}[\phi]$, Polchinski's ERG considers the
\textit{Wilson action}---a one-parameter family of action functionals
$\{S_{\Lambda^{\prime},\Lambda_{0}}[\phi]\}_{0\leq\Lambda^{\prime}\leq\Lambda_{0}}$
that depend on a lowered cutoff $\Lambda^{\prime}(\leq\Lambda_{0})$,
the original cutoff $\Lambda_{0}$ as well as the bare coupling
constants and satisfy the initial condition
$S_{\Lambda_{0},\Lambda_{0}}[\phi]=S_{\text{bare}}[\phi]$ at
$\Lambda^{\prime}=\Lambda_{0}$. Each Wilson action is chosen in such a
way that the correlation functions with respect to
$S_{\Lambda^{\prime},\Lambda_{0}}[\phi]$ coincide with the original
correlation functions with respect to $S_{\text{bare}}[\phi]$ if all
the external momenta are restricted below the lowered cutoff
$\Lambda^{\prime}$.\footnote{Strictly speaking, before taking the
  continuum limit, the correlation functions need not be the moments
  with respect to the canonical distribution; they can be replaced by
  objects that reduce to the moments below the cutoff scale. A primary
  example is Sonoda's \textit{modified correlation function}
  \cite{Sonoda:2015bla}, which is a normal-ordered correlation
  function with respect to a specific covariance that vanishes below
  the cutoff. Note that this generalization---one of the most general
  known formulations of the ERG---only alters the explicit form of the
  kernel, leaving the mapping structure \eqref{eq:16}
  intact. Therefore, our formalism is equally applicable to the ERG
  transformation \textit{\`{a} la} Sonoda.} A standard method for
constructing such a Wilson action is to integrate out the field lying
on the momentum shell $\Lambda^{\prime}\lesssim|p|\lesssim\Lambda_{0}$
in the functional-integral representation of the canonical partition
function. This procedure results in a linear map for Boltzmann weights
and is realized by a functional-integral transform that sends the
Boltzmann weight $\e^{-S_{\text{bare}}[\phi]}$ to another Boltzmann
weight $\e^{-S_{\Lambda^{\prime},\Lambda_{0}}[\phi]}$ (see appendix
\ref{appendix:A}). Since this transform can be applied repeatedly, one
can also construct a map from
$\e^{-S_{\Lambda^{\prime},\Lambda_{0}}[\phi]}$ to
$\e^{-S_{\Lambda,\Lambda_{0}}[\phi]}$ for
$\Lambda\leq\Lambda^{\prime}(\leq\Lambda_{0})$. We denote this map by
$\mathsf{R}_{\Lambda,\Lambda^{\prime}}$ and write
\begin{align}
  \mathsf{R}_{\Lambda,\Lambda^{\prime}}:\e^{-S_{\Lambda^{\prime},\Lambda_{0}}[\phi]}\mapsto\e^{-S_{\Lambda,\Lambda_{0}}[\phi]},\label{eq:15}
\end{align}
where
\begin{align}
  \e^{-S_{\Lambda,\Lambda_{0}}[\phi]}=\int_{Q}d\phi^{\prime}\,\mathsf{R}_{\Lambda,\Lambda^{\prime}}[\phi,\phi^{\prime}]\e^{-S_{\Lambda^{\prime},\Lambda_{0}}[\phi^{\prime}]}.\label{eq:16}
\end{align}
This is the ERG transformation, where
$\mathsf{R}_{\Lambda,\Lambda^{\prime}}[\ast,\ast]$ is the ERG kernel
and $\int_{Q}d\phi^{\prime}$ stands for the functional integral over
the configuration space $Q$. In this formulation, the continuum limit
refers to establishing the limit
$\lim_{\Lambda_{0}\to\infty}S_{\Lambda,\Lambda_{0}}[\phi]$ keeping
$\Lambda$ fixed. This limit could exist if the bare coupling constants
are appropriate functions of the original cutoff $\Lambda_{0}$, a
renormalization scale $\mu$, and a set of renormalized coupling
constants at the scale $\mu$. If such functions actually exist, the
theory is said to be exactly renormalizable. And the one-parameter
family of action functionals
$\{S_{\Lambda}[\phi]\}_{0\leq\Lambda<\infty}$ defined by
$S_{\Lambda}[\phi]=\lim_{\Lambda_{0}\to\infty}S_{\Lambda,\Lambda_{0}}[\phi]$
determines the correlation functions in the resultant continuum
theory; see, e.g., section 2.7 of the lecture notes
\cite{Sonoda:2007av}.

Now, the resemblance between \eqref{eq:5} and \eqref{eq:16} is
obvious. In this section, we will develop the general theory of the
ERG kernel on a multiply connected configuration space $Q$ under the
assumption that the ERG transformation is given by the
functional-integral transform \eqref{eq:16} for any $Q$. Just as in
the previous section, the keys are to lift the configuration space $Q$
to its universal covering space $\widetilde{Q}$ and then to introduce
an equivalence relation on the set of Boltzmann weights under the
action of the deck transformation group $\Gamma\cong\pi_{1}(Q)$. In
the following, we will first introduce the equivalence relation and
then construct the ERG kernel via the covering-space method. Several
additional properties will be imposed afterwards.

\subsection{Covering-space construction of ERG kernels}
\label{section:3.1}
Let us first introduce an equivalence relation on the set of Boltzmann
weights. Let $S_{1}$ and $S_{2}$ be two Euclidean actions of a field
$\phi$. The canonical distributions with respect to $S_{1}$ and
$S_{2}$ are said to be equivalent if the following equality holds for
all $\phi$:
\begin{align}
  \frac{\e^{-S_{1}[\phi]}}{Z_{1}}=\frac{\e^{-S_{2}[\phi]}}{Z_{2}},\label{eq:17}
\end{align}
or, equivalently,
\begin{align}
  \e^{-S_{1}[\phi]}=\frac{Z_{1}}{Z_{2}}\e^{-S_{2}[\phi]},\label{eq:18}
\end{align}
where $Z_{1}$ and $Z_{2}$ are the canonical partition functions with
respect to $S_{1}$ and $S_{2}$. Eq.~\eqref{eq:18} says that two
Boltzmann weights $\e^{-S_{1}[\phi]}$ and $\e^{-S_{2}[\phi]}$ give the
same canonical distribution if there exists a $\phi$-independent
nonzero (complex) constant $D$ such that the equality
$\e^{-S_{1}[\phi]}=D\e^{-S_{2}[\phi]}$ holds for all
$\phi$. Obviously, this defines an equivalence relation $\sim$ on the
set of Boltzmann weights; that is,
$\e^{-S_{1}[\phi]}\sim\e^{-S_{2}[\phi]}$ if
$\exists D\in\mathbb{C}^{\times}$
s.t. $\e^{-S_{1}[\phi]}=D\e^{-S_{2}[\phi]}$ for all
$\phi$.\footnote{$\mathbb{C}^{\times}=\{z\in\mathbb{C}\mid z\neq0\}$.}
This equivalence just says that there is always a constant-term
ambiguity in the definition of Euclidean action.

Now let us apply the above equivalence relation to Boltzmann weights
on a multiply connected configuration space
$Q\approx\widetilde{Q}/\Gamma$. First, let $\e^{-S[\phi]}$ be a
Boltzmann weight defined on the universal covering space
$\widetilde{Q}$, where $\phi\in\widetilde{Q}$. Given its universal
covering space, the configuration space $Q$ is constructed from
$\widetilde{Q}$ by identifying all the points
$\{\gamma\phi\}_{\gamma\in\Gamma}$ as a single point $\phi$, where
$\gamma\phi$ represents the action of a deck transformation
$\gamma\in\Gamma$ on $\phi\in\widetilde{Q}$. Correspondingly, all the
Boltzmann weights $\{\e^{-S[\gamma\phi]}\}_{\gamma\in\Gamma}$ for a
fixed $\phi$ should be equivalent to the Boltzmann weight
$\e^{-S[\phi]}$ for the same $\phi$; that is,
$\e^{-S[\gamma\phi]}\sim\e^{-S[\phi]}$. Thus we have
\begin{align}
  \e^{-S[\gamma\phi]}=D(\gamma)\e^{-S[\phi]},\label{eq:19}
\end{align}
where $D(\gamma)\in\mathbb{C}^{\times}$ is a $\phi$-independent
constant that may depend on $\gamma\in\Gamma$. Just as in section
\ref{section:2}, this constant cannot be arbitrary. Let $\gamma_{1}$
and $\gamma_{2}$ be two elements of $\Gamma$. Then, due to the group
closure property, the product $\gamma_{1}\gamma_{2}$ must also be an
element of $\Gamma$. Correspondingly, the following identity must hold
for an arbitrary Boltzmann weight:
\begin{align}
  \e^{-S[(\gamma_{1}\gamma_{2})\phi]}=\e^{-S[\gamma_{1}(\gamma_{2}\phi)]}.\label{eq:20}
\end{align}
It follows from \eqref{eq:19} that the left- and right-hand sides are
evaluated as
\begin{subequations}
  \begin{align}
    \text{LHS}&=\e^{-S[(\gamma_{1}\gamma_{2})\phi]}=D(\gamma_{1}\gamma_{2})\e^{-S[\phi]},\label{eq:21a}\\
    \text{RHS}&=\e^{-S[\gamma_{1}(\gamma_{2}\phi)]}=D(\gamma_{1})\e^{-S[\gamma_{2}\phi]}=D(\gamma_{1})D(\gamma_{2})\e^{-S[\phi]}.\label{eq:21b}
  \end{align}
\end{subequations}
Thus we find
$D(\gamma_{1}\gamma_{2})\e^{-S[\phi]}=D(\gamma_{1})D(\gamma_{2})\e^{-S[\phi]}$
for any $\e^{-S[\phi]}$, resulting in the following condition for any
$\gamma_{1},\gamma_{2}\in\Gamma$:
\begin{align}
  D(\gamma_{1}\gamma_{2})=D(\gamma_{1})D(\gamma_{2}).\label{eq:22}
\end{align}
Hence $D$ must be a map $D:\Gamma\to\mathbb{C}^{\times}$ that
preserves the group multiplication; that is, $D$ must be a
one-dimensional representation of
$\Gamma\cong\pi_{1}(Q)$. Eq.~\eqref{eq:19} corresponds to the twisted
boundary condition \eqref{eq:7} in quantum mechanics on $S^{1}$.

Let us next construct the ERG kernel on
$Q\approx\widetilde{Q}/\Gamma$. Let
$\widetilde{\mathsf{R}}_{\Lambda,\Lambda^{\prime}}[\ast,\ast]$ be the
ERG kernel on the universal covering space $\widetilde{Q}$ satisfying
\begin{align}
  \e^{-S_{\Lambda,\Lambda_{0}}[\phi]}=\int_{\widetilde{Q}}d\phi^{\prime}\,\widetilde{\mathsf{R}}_{\Lambda,\Lambda^{\prime}}[\phi,\phi^{\prime}]\e^{-S_{\Lambda^{\prime},\Lambda_{0}}[\phi^{\prime}]}.\label{eq:23}
\end{align}
Once given such a kernel, we can construct the ERG kernel on the
configuration space $Q\approx\widetilde{Q}/\Gamma$ by Dowker's
covering-space method. The key to this method is the following
identity:
\begin{align}
  \int_{\widetilde{Q}}d\phi\,F[\phi]=\int_{\widetilde{Q}/\Gamma}d\phi\sum_{\gamma\in\Gamma}F[\gamma\phi],\label{eq:24}
\end{align}
where $F$ is an arbitrary (integrable) functional. This identity just
says that first summing over the orbit
$\Gamma\phi=\{\gamma\phi\mid\gamma\in\Gamma\}$ of $\phi$ and then
integrating over the fundamental domain $Q\approx\widetilde{Q}/\Gamma$
of $\phi$ recovers the integration over the whole universal covering
space $\widetilde{Q}$. By using this, the functional integral
\eqref{eq:23} can be put into the following form:
\begin{align}
  \e^{-S_{\Lambda,\Lambda_{0}}[\phi]}
  &=\int_{\widetilde{Q}}d\phi^{\prime}\,\widetilde{\mathsf{R}}_{\Lambda,\Lambda^{\prime}}[\phi,\phi^{\prime}]\e^{-S_{\Lambda^{\prime},\Lambda_{0}}[\phi^{\prime}]}\nonumber\\
  &=\int_{\widetilde{Q}/\Gamma}d\phi^{\prime}\sum_{\gamma\in\Gamma}\widetilde{\mathsf{R}}_{\Lambda,\Lambda^{\prime}}[\phi,\gamma\phi^{\prime}]\e^{-S_{\Lambda^{\prime},\Lambda_{0}}[\gamma\phi^{\prime}]}\nonumber\\
  &=\int_{\widetilde{Q}/\Gamma}d\phi^{\prime}\sum_{\gamma\in\Gamma}\widetilde{\mathsf{R}}_{\Lambda,\Lambda^{\prime}}[\phi,\gamma\phi^{\prime}]D(\gamma)\e^{-S_{\Lambda^{\prime},\Lambda_{0}}[\phi^{\prime}]}\nonumber\\
  &=\int_{\widetilde{Q}/\Gamma}d\phi^{\prime}\left(\sum_{\gamma\in\Gamma}D(\gamma)\widetilde{\mathsf{R}}_{\Lambda,\Lambda^{\prime}}[\phi,\gamma\phi^{\prime}]\right)\e^{-S_{\Lambda^{\prime},\Lambda_{0}}[\phi^{\prime}]},\label{eq:25}
\end{align}
where the second equality follows from the identity \eqref{eq:24} and
the third equality follows from the twisted boundary condition
\eqref{eq:19}. Comparing \eqref{eq:25} with \eqref{eq:16}, we find the
following functional-integral kernel on
$Q\approx\widetilde{Q}/\Gamma$:
\begin{align}
  \mathsf{R}_{\Lambda,\Lambda^{\prime}}[\phi,\phi^{\prime}]=\sum_{\gamma\in\Gamma}D(\gamma)\widetilde{\mathsf{R}}_{\Lambda,\Lambda^{\prime}}[\phi,\gamma\phi^{\prime}].\label{eq:26}
\end{align}

We emphasize that \eqref{eq:26} is just the generic
functional-integral kernel acting on functionals that satisfy the
twisted boundary condition \eqref{eq:19}. To qualify as the ERG
kernel, \eqref{eq:26} must satisfy several additional properties. In
the following, we will impose three properties---the inhomogeneous
semigroup property, the reflection reality, and the normalization
condition---and clarify the additional conditions for the kernel
$\widetilde{\mathsf{R}}_{\Lambda,\Lambda^{\prime}}[\ast,\ast]$ as well
as the representation $D$.

\subsection{Property 1: Inhomogeneous semigroup property}
\label{section:3.2}
As already mentioned above \eqref{eq:15}, the ERG transformation can
be applied repeatedly; that is,
$\mathsf{R}_{\Lambda^{\prime},\Lambda^{\prime\prime}}$ followed by
$\mathsf{R}_{\Lambda,\Lambda^{\prime}}$ must result in
$\mathsf{R}_{\Lambda,\Lambda^{\prime\prime}}$. In addition,
$\mathsf{R}_{\Lambda,\Lambda}$ must leave the Boltzmann weight
unchanged. Hence the ERG transformation must form an
\textit{inhomogeneous semigroup} \cite{Plyushchev:1975}; that is, it
must satisfy the composition law
$\mathsf{R}_{\Lambda,\Lambda^{\prime}}\mathsf{R}_{\Lambda^{\prime},\Lambda^{\prime\prime}}=\mathsf{R}_{\Lambda,\Lambda^{\prime\prime}}$
and the initial condition $\mathsf{R}_{\Lambda,\Lambda}=\id$ for any
$\Lambda\leq\Lambda^{\prime}\leq\Lambda^{\prime\prime}(\leq\Lambda_{0})$,
where $\id$ stands for the identity operator. In terms of the kernel,
these algebraic properties are expressed as follows:
\begin{subequations}
  \begin{align}
    \text{(composition law)}&\quad\int_{Q}d\phi^{\prime}\,\mathsf{R}_{\Lambda,\Lambda^{\prime}}[\phi,\phi^{\prime}]\mathsf{R}_{\Lambda^{\prime},\Lambda^{\prime\prime}}[\phi^{\prime},\phi^{\prime\prime}]=\mathsf{R}_{\Lambda,\Lambda^{\prime\prime}}[\phi,\phi^{\prime\prime}],\label{eq:27a}\\
    \text{(initial condition)}&\quad\mathsf{R}_{\Lambda,\Lambda}[\phi,\phi^{\prime}]=\delta[\phi-\phi^{\prime}],\label{eq:27b}
  \end{align}
\end{subequations}
where $\phi,\phi^{\prime},\phi^{\prime\prime}\in Q$ and $\delta[\ast]$
stands for the delta-functional. These properties are guaranteed if
the kernel
$\widetilde{\mathsf{R}}_{\Lambda,\Lambda^{\prime}}[\ast,\ast]$
satisfies the following properties for arbitrary
$\phi,\phi^{\prime},\phi^{\prime\prime}\in\widetilde{Q}$ and
$\gamma\in\Gamma$:
\begin{subequations}
  \begin{align}
    \text{(composition law)}&\quad\int_{\widetilde{Q}}d\phi^{\prime}\,\widetilde{\mathsf{R}}_{\Lambda,\Lambda^{\prime}}[\phi,\phi^{\prime}]\widetilde{\mathsf{R}}_{\Lambda^{\prime},\Lambda^{\prime\prime}}[\phi^{\prime},\phi^{\prime\prime}]=\widetilde{\mathsf{R}}_{\Lambda,\Lambda^{\prime\prime}}[\phi,\phi^{\prime\prime}],\label{eq:28a}\\
    \text{(initial condition)}&\quad\widetilde{\mathsf{R}}_{\Lambda,\Lambda}[\phi,\phi^{\prime}]=\delta[\phi-\phi^{\prime}],\label{eq:28b}\\
    \text{($\Gamma$-invariance)}&\quad\widetilde{\mathsf{R}}_{\Lambda,\Lambda^{\prime}}[\gamma\phi,\gamma\phi^{\prime}]=\widetilde{\mathsf{R}}_{\Lambda,\Lambda^{\prime}}[\phi,\phi^{\prime}].\label{eq:28c}
  \end{align}
\end{subequations}
Under these assumptions, one can easily show that the ERG kernel
\eqref{eq:26} satisfies the desired properties. For instance, the
composition law \eqref{eq:27a} can be proved as follows:
\begin{align}
  &\int_{Q}d\phi^{\prime}\,\mathsf{R}_{\Lambda,\Lambda^{\prime}}[\phi,\phi^{\prime}]\mathsf{R}_{\Lambda^{\prime},\Lambda^{\prime\prime}}[\phi^{\prime},\phi^{\prime\prime}]\nonumber\\
  &=\int_{\widetilde{Q}/\Gamma}d\phi^{\prime}\left(\sum_{\gamma\in\Gamma}D(\gamma)\widetilde{\mathsf{R}}_{\Lambda,\Lambda^{\prime}}[\phi,\gamma\phi^{\prime}]\right)\left(\sum_{\gamma^{\prime}\in\Gamma}D(\gamma^{\prime})\widetilde{\mathsf{R}}_{\Lambda^{\prime},\Lambda^{\prime\prime}}[\phi^{\prime},\gamma^{\prime}\phi^{\prime\prime}]\right)\nonumber\\
  &=\int_{\widetilde{Q}/\Gamma}d\phi^{\prime}\sum_{\gamma\in\Gamma}\sum_{\gamma^{\prime}\in\Gamma}D(\gamma\gamma^{\prime})\widetilde{\mathsf{R}}_{\Lambda,\Lambda^{\prime}}[\phi,\gamma\phi^{\prime}]\widetilde{\mathsf{R}}_{\Lambda^{\prime},\Lambda^{\prime\prime}}[\gamma\phi^{\prime},\gamma\gamma^{\prime}\phi^{\prime\prime}]\nonumber\\
  &=\int_{\widetilde{Q}/\Gamma}d\phi^{\prime}\sum_{\gamma\in\Gamma}\sum_{\gamma^{\prime\prime}\in\Gamma}D(\gamma^{\prime\prime})\widetilde{\mathsf{R}}_{\Lambda,\Lambda^{\prime}}[\phi,\gamma\phi^{\prime}]\widetilde{\mathsf{R}}_{\Lambda^{\prime},\Lambda^{\prime\prime}}[\gamma\phi^{\prime},\gamma^{\prime\prime}\phi^{\prime\prime}]\nonumber\\
  &=\sum_{\gamma^{\prime\prime}\in\Gamma}D(\gamma^{\prime\prime})\int_{\widetilde{Q}/\Gamma}d\phi^{\prime}\sum_{\gamma\in\Gamma}\widetilde{\mathsf{R}}_{\Lambda,\Lambda^{\prime}}[\phi,\gamma\phi^{\prime}]\widetilde{\mathsf{R}}_{\Lambda^{\prime},\Lambda^{\prime\prime}}[\gamma\phi^{\prime},\gamma^{\prime\prime}\phi^{\prime\prime}]\nonumber\\
  &=\sum_{\gamma^{\prime\prime}\in\Gamma}D(\gamma^{\prime\prime})\int_{\widetilde{Q}}d\phi^{\prime}\,\widetilde{\mathsf{R}}_{\Lambda,\Lambda^{\prime}}[\phi,\phi^{\prime}]\widetilde{\mathsf{R}}_{\Lambda^{\prime},\Lambda^{\prime\prime}}[\phi^{\prime},\gamma^{\prime\prime}\phi^{\prime\prime}]\nonumber\\
  &=\sum_{\gamma^{\prime\prime}\in\Gamma}D(\gamma^{\prime\prime})\widetilde{\mathsf{R}}_{\Lambda,\Lambda^{\prime\prime}}[\phi,\gamma^{\prime\prime}\phi^{\prime\prime}]\nonumber\\
  &=\mathsf{R}_{\Lambda,\Lambda^{\prime\prime}}[\phi,\phi^{\prime\prime}],\label{eq:29}
\end{align}
where the first equality follows from $Q\approx\widetilde{Q}/\Gamma$
and \eqref{eq:26}, the second equality follows from the multiplication
rule $D(\gamma)D(\gamma^{\prime})=D(\gamma\gamma^{\prime})$ and the
$\Gamma$-invariance
$\widetilde{\mathsf{R}}_{\Lambda^{\prime},\Lambda^{\prime\prime}}[\phi^{\prime},\gamma^{\prime}\phi^{\prime\prime}]=\widetilde{\mathsf{R}}_{\Lambda^{\prime},\Lambda^{\prime\prime}}[\gamma\phi^{\prime},\gamma\gamma^{\prime}\phi^{\prime\prime}]$,
and the third equality follows from the change of summation variable
from $\gamma^{\prime}$ to
$\gamma^{\prime\prime}=\gamma\gamma^{\prime}$. The fifth equality
follows from the identity \eqref{eq:24}, the sixth equality follows
from the composition law \eqref{eq:28a}, and the last equality follows
from \eqref{eq:26}.

Similarly, the initial condition \eqref{eq:27b} is proved as follows:
\begin{align}
  \mathsf{R}_{\Lambda,\Lambda}[\phi,\phi^{\prime}]
  &=\sum_{\gamma\in\Gamma}D(\gamma)\widetilde{\mathsf{R}}_{\Lambda,\Lambda}[\phi,\gamma\phi^{\prime}]\nonumber\\
  &=\sum_{\gamma\in\Gamma}D(\gamma)\delta[\phi-\gamma\phi^{\prime}]\nonumber\\
  &=D(e)\delta[\phi-e\phi^{\prime}]\nonumber\\
  &=\delta[\phi-\phi^{\prime}],\label{eq:30}
\end{align}
where the second equality follows from the initial condition
\eqref{eq:28b}. The third equality follows from the fact that, for any
$\phi,\phi^{\prime}\in Q$, $\phi$ and $\gamma\phi^{\prime}$ never
become the same except for the case $\gamma=e$, where $e$ stands for
the identity element. This is just because $\gamma\phi^{\prime}$ lies
outside the fundamental domain $Q$ if $\gamma\neq e$. Finally, the
last equality follows from $D(e)=1$ for any one-dimensional
representation $D$ and $e\phi^{\prime}=\phi^{\prime}$ for any
$\phi^{\prime}$.

Summarizing, the sufficient condition for which the ERG kernel
\eqref{eq:26} satisfies the inhomogeneous semigroup properties
\eqref{eq:27a} and \eqref{eq:27b} is that the kernel
$\widetilde{\mathsf{R}}_{\Lambda,\Lambda^{\prime}}[\ast,\ast]$
fulfills the composition law \eqref{eq:28a}, the initial condition
\eqref{eq:28b}, and the $\Gamma$-invariance \eqref{eq:28c} on the
universal covering space $\widetilde{Q}$.

There are several key consequences that follow from the linear
combination \eqref{eq:26} and the algebraic properties
\eqref{eq:27a}--\eqref{eq:28c}. Among them are the following.
\begin{description}[leftmargin=0pt]
\item[Boundary condition.] Since the kernel
  $\widetilde{\mathsf{R}}_{\Lambda,\Lambda^{\prime}}[\ast,\ast]$ is
  defined on the universal covering space, the domain of
  $\mathsf{R}_{\Lambda,\Lambda^{\prime}}[\ast,\ast]$ defined by
  \eqref{eq:26} is naturally extended from $Q\times Q$ to
  $\widetilde{Q}\times\widetilde{Q}$. Under this extension, the ERG
  kernel satisfies the following twisted boundary conditions for any
  $\phi,\phi^{\prime}\in\widetilde{Q}$ and $\gamma\in\Gamma$:
  \begin{subequations}
    \begin{align}
      \mathsf{R}_{\Lambda,\Lambda^{\prime}}[\gamma\phi,\phi^{\prime}]&=D(\gamma)\mathsf{R}_{\Lambda,\Lambda^{\prime}}[\phi,\phi^{\prime}],\label{eq:31a}\\
      \mathsf{R}_{\Lambda,\Lambda^{\prime}}[\phi,\gamma\phi^{\prime}]&=\mathsf{R}_{\Lambda,\Lambda^{\prime}}[\phi,\phi^{\prime}]D(\gamma^{-1}).\label{eq:31b}
    \end{align}
  \end{subequations}
  Indeed, a straightforward calculation gives
  \begin{align}
    \mathsf{R}_{\Lambda,\Lambda^{\prime}}[\gamma\phi,\phi^{\prime}]
    &=\sum_{\gamma^{\prime}\in\Gamma}D(\gamma^{\prime})\widetilde{\mathsf{R}}_{\Lambda,\Lambda^{\prime}}[\gamma\phi,\gamma^{\prime}\phi^{\prime}]\nonumber\\
    &=\sum_{\gamma^{\prime}\in\Gamma}D(\gamma\gamma^{-1}\gamma^{\prime})\widetilde{\mathsf{R}}_{\Lambda,\Lambda^{\prime}}[\phi,\gamma^{-1}\gamma^{\prime}\phi^{\prime}]\nonumber\\
    &=D(\gamma)\sum_{\gamma^{\prime\prime}\in\Gamma}D(\gamma^{\prime\prime})\widetilde{\mathsf{R}}_{\Lambda,\Lambda^{\prime}}[\phi,\gamma^{\prime\prime}\phi^{\prime}]\nonumber\\
    &=D(\gamma)\mathsf{R}_{\Lambda,\Lambda^{\prime}}[\phi,\phi^{\prime}],\label{eq:32}
  \end{align}
  where the second equality follows from the $\Gamma$-invariance
  $\widetilde{\mathsf{R}}_{\Lambda,\Lambda^{\prime}}[\gamma\phi,\gamma^{\prime}\phi^{\prime}]=\widetilde{\mathsf{R}}_{\Lambda,\Lambda^{\prime}}[\gamma^{-1}\gamma\phi,\gamma^{-1}\gamma^{\prime}\phi^{\prime}]$
  and the third equality follows from the change of summation variable
  from $\gamma^{\prime}$ to
  $\gamma^{\prime\prime}=\gamma^{-1}\gamma^{\prime}$. Similarly, one
  can prove \eqref{eq:31b} under the assumption of the
  $\Gamma$-invariance \eqref{eq:28c}. Importantly, the twisted
  boundary condition for the Boltzmann weights can be reproduced from
  \eqref{eq:31a}. By using \eqref{eq:16} and \eqref{eq:31a}, we have
  \begin{align}
    \e^{-S_{\Lambda,\Lambda_{0}}[\gamma\phi]}
    &=\int_{Q}d\phi^{\prime}\,\mathsf{R}_{\Lambda,\Lambda^{\prime}}[\gamma\phi,\phi^{\prime}]\e^{-S_{\Lambda^{\prime},\Lambda_{0}}[\phi^{\prime}]}\nonumber\\
    &=D(\gamma)\int_{Q}d\phi^{\prime}\,\mathsf{R}_{\Lambda,\Lambda^{\prime}}[\phi,\phi^{\prime}]\e^{-S_{\Lambda^{\prime},\Lambda_{0}}[\phi^{\prime}]}\nonumber\\
    &=D(\gamma)\e^{-S_{\Lambda,\Lambda_{0}}[\phi]}\label{eq:33}
  \end{align}
  for any $\phi\in\widetilde{Q}$ and $\gamma\in\Gamma$.
\item[ERG flow equation.] Since the ERG transformation
  $\mathsf{R}_{\Lambda,\Lambda^{\prime}}$ forms a continuous
  inhomogeneous semigroup, it has the generator $\mathsf{G}_{\Lambda}$
  defined by
  $\mathsf{G}_{\Lambda}=\Lambda\lim_{\delta\Lambda\to0}(\mathsf{R}_{\Lambda-\delta\Lambda,\Lambda}-\mathsf{R}_{\Lambda,\Lambda})/\delta\Lambda$. (Here
  we are considering the derivative along the direction of decreasing
  $\Lambda$.) Let $\mathsf{G}_{\Lambda}[\phi,\delta/\delta\phi]$ be a
  functional-differential-operator realization of the ERG generator on
  the universal covering space; that is, let
  $\widetilde{\mathsf{R}}_{\Lambda,\Lambda^{\prime}}[\ast,\ast]$
  satisfy the following local functional-differential equation (flow
  equation):
  \begin{align}
    -\Lambda\frac{\partial}{\partial\Lambda}\widetilde{\mathsf{R}}_{\Lambda,\Lambda^{\prime}}[\phi,\phi^{\prime}]=\mathsf{G}_{\Lambda}[\phi,\tfrac{\delta}{\delta\phi}]\,\widetilde{\mathsf{R}}_{\Lambda,\Lambda^{\prime}}[\phi,\phi^{\prime}].\label{eq:34}
  \end{align}
  It then follows from the linear combination \eqref{eq:26} that the
  ERG kernel $\mathsf{R}_{\Lambda,\Lambda^{\prime}}[\ast,\ast]$
  satisfies the same flow equation as \eqref{eq:34}:
  \begin{align}
    -\Lambda\frac{\partial}{\partial\Lambda}\mathsf{R}_{\Lambda,\Lambda^{\prime}}[\phi,\phi^{\prime}]=\mathsf{G}_{\Lambda}[\phi,\tfrac{\delta}{\delta\phi}]\,\mathsf{R}_{\Lambda,\Lambda^{\prime}}[\phi,\phi^{\prime}].\label{eq:35}
  \end{align}
  It then follows from the linear functional-integral transform
  \eqref{eq:16} that the Boltzmann weight
  $\e^{-S_{\Lambda,\Lambda_{0}}[\phi]}$ also satisfies the same flow
  equation as \eqref{eq:34}:
  \begin{align}
    -\Lambda\frac{\partial}{\partial\Lambda}\e^{-S_{\Lambda,\Lambda_{0}}[\phi]}=\mathsf{G}_{\Lambda}[\phi,\tfrac{\delta}{\delta\phi}]\e^{-S_{\Lambda,\Lambda_{0}}[\phi]}.\label{eq:36}
  \end{align}
  An important point to note here is that, though all these objects
  satisfy the same flow equation, their boundary conditions are
  different. For instance,
  $\mathsf{R}_{\Lambda,\Lambda^{\prime}}[\ast,\ast]$ satisfies the
  twisted boundary conditions \eqref{eq:31a} and \eqref{eq:31b},
  whereas
  $\widetilde{\mathsf{R}}_{\Lambda,\Lambda^{\prime}}[\ast,\ast]$
  satisfies the $\Gamma$-invariance \eqref{eq:28c}. In other words,
  without specifying the correct boundary condition, the flow equation
  cannot describe the global topology of $Q$. This is as expected
  because $Q$ and $\widetilde{Q}$ are locally the same. Note, however,
  that there exists an alternative gauge-equivalent flow equation that
  captures the topological effect at the level of local
  functional-differential equation, where the functional derivative
  $\delta/\delta\phi$ is replaced by a covariant derivative
  $\delta/\delta\phi+i\mathscr{A}$ with respect to a background
  Abelian gauge field $\mathscr{A}$; see section \ref{section:4}.
\item[Nonrenormalization of weight factor.] So far we have implicitly
  assumed that all the weight factors are independent of the lowered
  cutoff $\Lambda$. This assumption can be easily verified as
  follows. Suppose that the Boltzmann weight
  $\e^{-S_{\Lambda,\Lambda_{0}}[\phi]}$ satisfies the following
  twisted boundary condition for any $\Lambda(\leq\Lambda_{0})$ and
  $\gamma\in\Gamma$:
  \begin{align}
    \e^{-S_{\Lambda,\Lambda_{0}}[\gamma\phi]}=D_{\Lambda,\Lambda_{0}}(\gamma)\e^{-S_{\Lambda,\Lambda_{0}}[\phi]},\label{eq:37}
  \end{align}
  where $D_{\Lambda,\Lambda_{0}}(\gamma)$ depends on both $\Lambda$
  and $\Lambda_{0}$. In this case, by substituting
  $\e^{-S_{\Lambda^{\prime},\Lambda_{0}}[\gamma\phi]}=D_{\Lambda^{\prime},\Lambda_{0}}(\gamma)\e^{-S_{\Lambda^{\prime},\Lambda_{0}}[\phi]}$
  into the third line of \eqref{eq:25}, we get
  \begin{align}
    \mathsf{R}_{\Lambda,\Lambda^{\prime}}[\phi,\phi^{\prime}]=\sum_{\gamma\in\Gamma}D_{\Lambda^{\prime},\Lambda_{0}}(\gamma)\widetilde{\mathsf{R}}_{\Lambda,\Lambda^{\prime}}[\phi,\gamma\phi^{\prime}].\label{eq:38}
  \end{align}
  It then follows from \eqref{eq:32} that this kernel satisfies the
  following twisted boundary condition:
  \begin{align}
    \mathsf{R}_{\Lambda,\Lambda^{\prime}}[\gamma\phi,\phi^{\prime}]=D_{\Lambda^{\prime},\Lambda_{0}}(\gamma)\mathsf{R}_{\Lambda,\Lambda^{\prime}}[\phi,\phi^{\prime}].\label{eq:39}
  \end{align}
  However, it follows from \eqref{eq:33} that \eqref{eq:39} induces
  the following twisted boundary condition for the Boltzmann weight
  $\e^{-S_{\Lambda,\Lambda_{0}}[\phi]}$:
  \begin{align}
    \e^{-S_{\Lambda,\Lambda_{0}}[\gamma\phi]}=D_{\Lambda^{\prime},\Lambda_{0}}(\gamma)\e^{-S_{\Lambda,\Lambda_{0}}[\phi]}.\label{eq:40}
  \end{align}
  Obviously, eqs.~\eqref{eq:37} and \eqref{eq:40} are compatible with
  each other if and only if the following equality holds for any
  $\Lambda,\Lambda^{\prime}(\leq\Lambda_{0})$:
  \begin{align}
    D_{\Lambda,\Lambda_{0}}(\gamma)=D_{\Lambda^{\prime},\Lambda_{0}}(\gamma).\label{eq:41}
  \end{align}
  This equation says that the weight factor does not depend on the
  lowered cutoff. Indeed, by choosing $\Lambda^{\prime}=\Lambda_{0}$
  in \eqref{eq:41}, we have
  $D_{\Lambda,\Lambda_{0}}(\gamma)=D_{\Lambda_{0},\Lambda_{0}}(\gamma)$
  for any $\Lambda$, meaning that $D_{\Lambda,\Lambda_{0}}(\gamma)$ is
  independent of $\Lambda$. Put differently, the weight factor never
  undergoes renormalization under the ERG. Thus we arrive at the
  following:
  \begin{theorem*}[Nonrenormalization of weight factor] The
    one-dimensional representation $D:\Gamma\to\mathbb{C}^{\times}$
    appearing in the ERG kernel \eqref{eq:26} is independent of the
    lowered cutoff $\Lambda(<\Lambda_{0})$ and satisfies
    \begin{align}
      \Lambda\frac{\partial}{\partial\Lambda}D(\gamma)=0,\quad\forall\gamma\in\Gamma\cong\pi_{1}(Q).\label{eq:42}
    \end{align}
    In other words, the weight factor is never renormalized under the
    ERG.
  \end{theorem*}
  As we will see in section \ref{section:4}, this provides a simple
  nonperturbative proof for the nonrenormalization of
  Wess-Zumino-Witten terms in nonlinear sigma models.
\end{description}

\subsection{Property 2: Reflection reality}
\label{section:3.3}
As is well known, Euclidean action can be complex in the presence of
topological terms. A typical example is the four-dimensional
Yang-Mills gauge theory in the presence of the $\theta$-term, which
breaks time-reversal invariance. More generally, Euclidean action
becomes complex if the action is not invariant under reflection, or
orientation-reversal of Euclidean spacetime $X$. Even in such a
theory, however, Euclidean action becomes invariant under the
composite operation of complex conjugation and orientation
reversal. This invariance---the \textit{reflection reality}---is the
reality property for Euclidean action that is needed to guarantee the
reflection positivity of the Osterwalder-Schrader axioms. In this
section, we will clarify the condition for the kernel
$\widetilde{\mathsf{R}}_{\Lambda,\Lambda^{\prime}}[\ast,\ast]$ and the
representation $D$ for which the Wilson action satisfies the
reflection reality. In the following, we will assume that the
Euclidean spacetime $X$ is an orientable manifold and the
configuration space $Q$ is the based mapping space \eqref{eq:1}. A
generalization to the unbased mapping space is straightforward.

\begin{figure}[t]
  \centering%
  \input{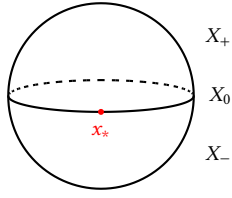}%
  \caption{Cutting the Euclidean spacetime $X$ in half along a
    codimension-one hypersurface $X_{0}$. In this figure, $X$ and
    $X_{0}$ are represented as a sphere and its equator. The red dot
    represents the base point $x_{\ast}\in X_{0}$. $\Phi[\phi]$ and
    $\Psi[\vartheta\phi]$ in the Osterwalder-Schrader inner product
    \eqref{eq:44} reside only on the northern hemisphere $X_{+}$ and
    the southern hemisphere $X_{-}$, respectively.}
  \label{figure:1}
\end{figure}

To begin with, let us first recall the reflection positivity in
Euclidean field theory. We first cut the Euclidean spacetime $X$ along
a codimension-one hypersurface $X_{0}$ that contains the base point
$x_{\ast}$ (see figure \ref{figure:1}). This splits the spacetime $X$
into three parts---the future part $X_{+}$, the past part $X_{-}$, and
the hypersurface $X_{0}$, where $X_{+}$ and $X_{-}$ have the same
boundary $X_{0}$ and are transformed to each other by the reflection
$r:X_{\pm}\to X_{\mp}$ across the hypersurface $X_{0}$. Note that the
reflection $r$ is an involution and leaves the base point
$x_{\ast}\in X_{0}$ unchanged; that is, it satisfies $r^{2}=\id$ and
$rx_{\ast}=x_{\ast}$. Let $\Psi[\phi]$ be an arbitrary
(square-integrable) functional of $\phi$ whose support is
$X_{+}$. Then, the reflection positivity refers to the following
inequality (see, e.g., \cite{Glimm:1987ylb}):
\begin{align}
  (\Psi,\Psi)_{\text{OS}}\geq0,\label{eq:43}
\end{align}
where $(\ast,\ast)_{\text{OS}}$ stands for the Osterwalder-Schrader
inner product defined by
\begin{align}
  (\Psi,\Phi)_{\text{OS}}=\int_{Q}d\phi\,\overline{\Psi[\vartheta\phi]}\Phi[\phi]\frac{\e^{-S[\phi]}}{Z}.\label{eq:44}
\end{align}
Here the overline ($\overline{\phantom{m}}$) denotes the complex
conjugate. $\vartheta:Q\to Q$ is a base-point-preserving involution
induced by the reflection $r$; that is, it is defined by
$(\vartheta\phi)(x)=\phi(rx)$, or, equivalently,\footnote{The
  definition $(\vartheta\phi)(x)=\phi(rx)$ says that the map
  $\vartheta\phi:X\to Y$ is defined by the composed map
  $X\xrightarrow{r}X\xrightarrow{\phi}Y$.}
\begin{align}
  \vartheta\phi=\phi\circ r,\label{eq:45}
\end{align}
where $\circ$ stands for the composition of maps. Clearly, the
so-defined map satisfies the involution property
$\vartheta^{2}\phi=\vartheta(\vartheta\phi)=\vartheta(\phi\circ
r)=(\phi\circ r)\circ r=\phi\circ r^{2}=\phi$ as well as the
base-point-preserving property
$(\vartheta\phi)(x_{\ast})=\phi(rx_{\ast})=\phi(x_{\ast})=y_{\ast}$
for any $\phi\in Q$. In the following, we will also call $\vartheta$
the reflection.

Now, to fulfill the inequality, the left-hand side of \eqref{eq:43}
must first be real; that is, it must satisfy\footnote{One can also
  arrive at the same result \eqref{eq:48} by imposing the stronger
  condition
  $(\Psi,\Phi)_{\text{OS}}=\overline{(\Phi,\Psi)_{\text{OS}}}$.}
\begin{align}
  (\Psi,\Psi)_{\text{OS}}=\overline{(\Psi,\Psi)_{\text{OS}}}.\label{eq:46}
\end{align}
This gives a nontrivial constraint for the Euclidean action. The
right-hand side can be written as
\begin{align}
  \overline{(\Psi,\Psi)_{\text{OS}}}
  &=\overline{\int_{Q}d\phi^{\prime}\,\overline{\Psi[\vartheta\phi^{\prime}]}\Psi[\phi^{\prime}]\,\frac{\e^{-S[\phi^{\prime}]}}{Z}}\nonumber\\
  &=\int_{Q}d\phi^{\prime}\,\Psi[\vartheta\phi^{\prime}]\overline{\Psi[\phi^{\prime}]}\,\frac{\e^{-\overline{S[\phi^{\prime}]}}}{\overline{Z}}\nonumber\\
  &=\int_{Q}d\phi\,\Psi[\phi]\overline{\Psi[\vartheta\phi]}\,\frac{\e^{-\overline{S[\vartheta\phi]}}}{\overline{Z}},\label{eq:47}
\end{align}
where in the second line we have assumed that the volume element
$d\phi^{\prime}$ is real. In the last line, we have changed the
integration variable from $\phi^{\prime}$ to
$\phi=\vartheta\phi^{\prime}$ and assumed that the volume element is
invariant under this change of variable, $d\phi^{\prime}=d\phi$.
Comparing \eqref{eq:47} with \eqref{eq:44} for $\Phi=\Psi$, we obtain
\begin{align}
  \frac{\e^{-\overline{S[\vartheta\phi]}}}{\overline{Z}}=\frac{\e^{-S[\phi]}}{Z},\label{eq:48}
\end{align}
which is satisfied if the Euclidean action fulfills the following
equality:
\begin{align}
  \overline{S[\vartheta\phi]}=S[\phi].\label{eq:49}
\end{align}
This is the reflection reality of Euclidean action.

Next, to understand the reflection reality for
$\widetilde{\mathsf{R}}_{\Lambda,\Lambda^{\prime}}[\ast,\ast]$ and
$D$, we have to introduce the reflection $\widetilde{\vartheta}$ on
the universal covering space $\widetilde{Q}$ and uncover its action on
the deck transformation group $\Gamma$. This is most easily done by
representing $\widetilde{Q}$ as homotopy classes of paths on $Q$. Let
us begin with some definitions.

First, a path on the based mapping space
$Q=\Map_{\ast}(X,Y)=\{\phi:X\to Y\mid\phi(x_{\ast})=y_{\ast}\}$ is a
one-parameter family of fields that satisfy the boundary condition
$\phi(x_{\ast})=y_{\ast}$. More precisely, a path $\ell$ on $Q$ is a
map
\begin{align}
  \ell:[0,1]\to Q\label{eq:50}
\end{align}
that satisfies the base-point-preserving property
$\ell(t)(x_{\ast})=y_{\ast}$ for any $t\in[0,1]$. Given such a path
$\ell$, we call the map $\ell^{-1}:[0,1]\to Q$ defined by
$\ell^{-1}(t)=\ell(1-t)$ the inverse path of $\ell$.

Let $\ell_{1},\ell_{2}:[0,1]\to Q$ be two paths satisfying
$\ell_{1}(1)=\ell_{2}(0)$. Then we can define the product path
$\ell_{1}\ast\ell_{2}:[0,1]\to Q$ by gluing the ending point of
$\ell_{1}$ with the starting point of $\ell_{2}$:
\begin{align}
  (\ell_{1}\ast\ell_{2})(t)=
  \begin{cases}
    \ell_{1}(2t)&0\leq t\leq\frac{1}{2},\\
    \ell_{2}(2t-1)&\frac{1}{2}\leq t\leq1.\\
  \end{cases}\label{eq:51}
\end{align}

Let us next consider the reflection of a path $\ell:[0,1]\to Q$. Since
$\ell(t)$ is an element of $Q$ for each $t\in[0,1]$, we can define the
reflection $\ell\mapsto\vartheta\ell$ in exactly the same way as
\eqref{eq:45}:
\begin{align}
  (\vartheta\ell)(t)=\ell(t)\circ r.\label{eq:52}
\end{align}
We will call the so-defined path $\vartheta\ell:[0,1]\to Q$ the
\textit{reflected path}. It follows from \eqref{eq:51} and
\eqref{eq:52} that the reflection $\vartheta$ acts on the product path
$\ell_{1}\ast\ell_{2}$ as
$\vartheta(\ell_{1}\ast\ell_{2})=\vartheta\ell_{1}\ast\vartheta\ell_{2}$.

Now, it is well known in algebraic topology that the universal
covering space of $Q$ can be represented as the set of homotopy
classes of paths starting from a specifically chosen base point on
$Q$; see, e.g., Section 1.3 of Hatcher's textbook
\cite{Hatcher:2001}. As a base point on $Q$, we choose the
base-point-preserving constant map (i.e., the constant field)
$\phi_{\ast}\in Q$ satisfying $\phi_{\ast}(x)=y_{\ast}$ for any
$x\in X$. In this case, the universal covering space $\widetilde{Q}$
can be represented as
\begin{align}
  \widetilde{Q}=\{[\ell^{\prime}]\mid\ell^{\prime}:[0,1]\to Q~~\&~~\ell^{\prime}(0)=\phi_{\ast}\},\label{eq:53}
\end{align}
where $[\ell^{\prime}]$ stands for the homotopy class of the path
$\ell^{\prime}$ with respect to homotopies that fix the endpoints
$\ell^{\prime}(0)$ and $\ell^{\prime}(1)$. Let $\ell$ be an arbitrary
closed path satisfying $\ell(0)=\ell(1)=\phi_{\ast}$. Then we can
define the map $\gamma_{[\ell]}:\widetilde{Q}\to\widetilde{Q}$ by
\begin{align}
  \gamma_{[\ell]}[\ell^{\prime}]=[\ell\ast\ell^{\prime}].\label{eq:54}
\end{align}
This is the deck transformation associated with the homotopy class
$[\ell]$ of the closed path $\ell$. Clearly, the so-defined
transformation satisfies the composition law
$\gamma_{[\ell_{1}]}\gamma_{[\ell_{2}]}=\gamma_{[\ell_{1}\ast\ell_{2}]}$
for any closed paths $\ell_{1}$ and $\ell_{2}$ satisfying
$\ell_{1}(0)=\ell_{1}(1)=\ell_{2}(0)=\ell_{2}(1)=\phi_{\ast}$. It is
also clear that $\gamma_{[\ell^{-1}]}=\gamma_{[\ell]}^{-1}$ for any
closed path $\ell$ based at $\phi_{\ast}$.

Let us next introduce the reflection $\widetilde{\vartheta}$ on the
universal covering space $\widetilde{Q}$. This can be done by defining
the map $\widetilde{\vartheta}:\widetilde{Q}\to\widetilde{Q}$ by
\begin{align}
  \widetilde{\vartheta}[\ell^{\prime}]=[\vartheta\ell^{\prime}].\label{eq:55}
\end{align}
Indeed, the so-defined map satisfies the involution property
$\widetilde{\vartheta}^{2}[\ell^{\prime}]=\widetilde{\vartheta}[\vartheta\ell^{\prime}]=[\vartheta^{2}\ell^{\prime}]=[\ell^{\prime}]$
for any $[\ell^{\prime}]\in\widetilde{Q}$. It is now easy to show that
the conjugation
$\widetilde{\vartheta}\gamma_{[\ell]}\widetilde{\vartheta}$ acts on
$\widetilde{Q}$ as follows:
\begin{align}
  \widetilde{\vartheta}\gamma_{[\ell]}\widetilde{\vartheta}[\ell^{\prime}]
  &=\widetilde{\vartheta}\gamma_{[\ell]}[\vartheta\ell^{\prime}]\nonumber\\
  &=\widetilde{\vartheta}[\ell\ast\vartheta\ell^{\prime}]\nonumber\\
  &=[\vartheta(\ell\ast\vartheta\ell^{\prime})]\nonumber\\
  &=[\vartheta\ell\ast\vartheta^{2}\ell^{\prime}]\nonumber\\
  &=[\vartheta\ell\ast\ell^{\prime}]\nonumber\\
  &=\gamma_{[\vartheta\ell]}[\ell^{\prime}],\label{eq:56}
\end{align}
where we have used \eqref{eq:55}, \eqref{eq:54},
$\vartheta(\ell_{1}\ast\ell_{2})=\vartheta\ell_{1}\ast\vartheta\ell_{2}$,
and $\vartheta^{2}=\id$. Thus we obtain the following formula:
\begin{align}
  \widetilde{\vartheta}\gamma_{[\ell]}\widetilde{\vartheta}=\gamma_{[\vartheta\ell]}.\label{eq:57}
\end{align}
This formula plays an important role in the reflection reality of
weight factors.

Now we are ready to study additional conditions for the kernel
$\widetilde{\mathsf{R}}_{\Lambda,\Lambda^{\prime}}[\ast,\ast]$ and the
representation $D$. Let us first consider the constraint on the
representation $D$ by combining the reflection reality and the twisted
boundary condition. Let $\e^{-S[\phi]}$ be a Boltzmann weight on the
universal covering space $\widetilde{Q}$ satisfying the reflection
reality $\overline{\e^{-S[\widetilde{\vartheta}\phi]}}=\e^{-S[\phi]}$
for any $\phi\in\widetilde{Q}$. Then, by just replacing $\phi$ with
$\gamma_{[\ell]}\phi\in\widetilde{Q}$, the following equality must
hold for any $\gamma_{[\ell]}\in\Gamma$:
\begin{align}
  \overline{\e^{-S[\widetilde{\vartheta}(\gamma_{[\ell]}\phi)]}}=\e^{-S[\gamma_{[\ell]}\phi]}.\label{eq:58}
\end{align}
By using the twisted boundary condition \eqref{eq:19}, the left- and
right-hand sides are evaluated as follows:
\begin{subequations}
  \begin{align}
    \text{LHS}&=\overline{\e^{-S[\widetilde{\vartheta}(\gamma_{[\ell]}\phi)]}}=\overline{\e^{-S[(\widetilde{\vartheta}\gamma_{[\ell]}\widetilde{\vartheta})(\widetilde{\vartheta}\phi)]}}=\overline{\e^{-S[\gamma_{[\vartheta\ell]}(\widetilde{\vartheta}\phi)]}}=\overline{D(\gamma_{[\vartheta\ell]})\e^{-S[\widetilde{\vartheta}\phi]}}=\overline{D(\gamma_{[\vartheta\ell]})}\e^{-S[\phi]},\label{eq:59a}\\
    \text{RHS}&=\e^{-S[\gamma_{[\ell]}\phi]}=D(\gamma_{[\ell]})\e^{-S[\phi]},\label{eq:59b}
  \end{align}
\end{subequations}
where the second and third equalities of \eqref{eq:59a} follow from
$\widetilde{\vartheta}^{2}=\id$ and \eqref{eq:57}. Comparing
\eqref{eq:59a} and \eqref{eq:59b}, we obtain
\begin{align}
  \overline{D(\gamma_{[\vartheta\ell]})}=D(\gamma_{[\ell]})\label{eq:60}
\end{align}
for any $\gamma_{[\ell]}\in\Gamma$. This is the reflection reality for
the one-dimensional representation of the deck transformation
group. This equation says that the weight factors must be invariant
under the composite operation of complex conjugation and reflection of
closed paths.

Let us next study the constraint on the kernel
$\widetilde{\mathsf{R}}_{\Lambda,\Lambda^{\prime}}[\ast,\ast]$ by
studying the ERG transformation. Let $S_{\Lambda,\Lambda_{0}}[\phi]$
be a Wilson action satisfying the reflection reality
$\overline{\e^{-S_{\Lambda,\Lambda_{0}}[\vartheta\phi]}}=\e^{-S_{\Lambda,\Lambda_{0}}[\phi]}$
for any $\Lambda\leq\Lambda_{0}$ and $\phi\in Q$. By using the
functional-integral transform \eqref{eq:16}, the left-hand side can be
rewritten as follows:
\begin{align}
  \overline{\e^{-S_{\Lambda,\Lambda_{0}}[\vartheta\phi]}}
  &=\overline{\int_{Q}d\phi^{\prime\prime}\,\mathsf{R}_{\Lambda,\Lambda^{\prime}}[\vartheta\phi,\phi^{\prime\prime}]\e^{-S_{\Lambda^{\prime},\Lambda_{0}}[\phi^{\prime\prime}]}}\nonumber\\
  &=\int_{Q}d\phi^{\prime\prime}\,\overline{\mathsf{R}_{\Lambda,\Lambda^{\prime}}[\vartheta\phi,\phi^{\prime\prime}]}\,\overline{\e^{-S_{\Lambda^{\prime},\Lambda_{0}}[\phi^{\prime\prime}]}}\nonumber\\
  &=\int_{Q}d\phi^{\prime}\,\overline{\mathsf{R}_{\Lambda,\Lambda^{\prime}}[\vartheta\phi,\vartheta\phi^{\prime}]}\,\overline{\e^{-S_{\Lambda^{\prime},\Lambda_{0}}[\vartheta\phi^{\prime}]}}\nonumber\\
  &=\int_{Q}d\phi^{\prime}\,\overline{\mathsf{R}_{\Lambda,\Lambda^{\prime}}[\vartheta\phi,\vartheta\phi^{\prime}]}\e^{-S_{\Lambda^{\prime},\Lambda_{0}}[\phi^{\prime}]},\label{eq:61}
\end{align}
where in the third equality we have changed the integration variable
from $\phi^{\prime\prime}$ to
$\phi^{\prime}=\vartheta\phi^{\prime\prime}$ and assumed that the
volume element satisfies $d\phi^{\prime\prime}=d\phi^{\prime}$. The
last equality of \eqref{eq:61} follows from the reflection reality
$\overline{\e^{-S_{\Lambda^{\prime},\Lambda_{0}}[\vartheta\phi]}}=\e^{-S_{\Lambda^{\prime},\Lambda_{0}}[\phi]}$. Comparing
the last line of \eqref{eq:61} with \eqref{eq:16}, we get
\begin{align}
  \overline{\mathsf{R}_{\Lambda,\Lambda^{\prime}}[\vartheta\phi,\vartheta\phi^{\prime}]}=\mathsf{R}_{\Lambda,\Lambda^{\prime}}[\phi,\phi^{\prime}].\label{eq:62}
\end{align}
This is the reflection reality for the ERG kernel. Similarly, by
applying the same argument for the functional-integral transform
\eqref{eq:23} on the universal covering space $\widetilde{Q}$, we
obtain the following condition for the kernel
$\widetilde{\mathsf{R}}_{\Lambda,\Lambda^{\prime}}[\ast,\ast]$:
\begin{align}
  \overline{\widetilde{\mathsf{R}}_{\Lambda,\Lambda^{\prime}}[\widetilde{\vartheta}\phi,\widetilde{\vartheta}\phi^{\prime}]}=\widetilde{\mathsf{R}}_{\Lambda,\Lambda^{\prime}}[\phi,\phi^{\prime}].\label{eq:63}
\end{align}
One can easily show that the linear combination \eqref{eq:26}
satisfies the reflection reality \eqref{eq:62} if the kernel
$\widetilde{\mathsf{R}}_{\Lambda,\Lambda^{\prime}}[\ast,\ast]$ and the
representation $D$ fulfill the conditions \eqref{eq:63} and
\eqref{eq:60}.

Now, as we will see in section \ref{section:4}, the one-dimensional
representation $D$ can be represented as a Wilson loop
$D(\gamma_{[\ell]})=\exp(i\oint_{\ell}\mathscr{A})$ of a background
gauge field $\mathscr{A}$ along a closed path $\ell$. If the gauge
field is a pseudo-tensor (i.e., it changes its sign under the
orientation reversal), the Wilson loop along the reflected path
$\vartheta\ell$ becomes equivalent to the Wilson loop along the
inverse path $\ell^{-1}$. Namely,
\begin{align}
  D(\gamma_{[\vartheta\ell]})=D(\gamma_{[\ell]}^{-1})\quad\text{if $\mathscr{A}$ is a pseudo-tensor}.\label{eq:64}
\end{align}
Combining this with the reflection reality \eqref{eq:60}, we have
$\overline{D(\gamma_{[\ell]}^{-1})}=D(\gamma_{[\ell]})$; that is,
$\overline{D(\gamma)}=D(\gamma^{-1})$ for any $\gamma\in\Gamma$. In
other words, $D$ becomes a one-dimensional unitary representation
$D:\Gamma\to U(1)$, because
$|D(\gamma)|^{2}=\overline{D(\gamma)}D(\gamma)=D(\gamma^{-1})D(\gamma)=D(\gamma^{-1}\gamma)=D(e)=1$
for any $\gamma\in\Gamma$. We will see in section \ref{section:4} that
such a unitary representation of $\Gamma$ is related to
Wess-Zumino-Witten terms in nonlinear sigma models.

\subsection{Property 3: Normalization condition}
\label{section:3.4}
The ERG transformation satisfying the composition law
$\mathsf{R}_{\Lambda,\Lambda^{\prime}}\mathsf{R}_{\Lambda^{\prime},\Lambda^{\prime\prime}}=\mathsf{R}_{\Lambda,\Lambda^{\prime\prime}}$
and the initial condition $\mathsf{R}_{\Lambda,\Lambda}=\id$ has a
normalization ambiguity. In fact, by redefining
$\mathsf{R}_{\Lambda,\Lambda^{\prime}}$ as
$\mathsf{R}_{\Lambda,\Lambda^{\prime}}^{\prime}=a_{\Lambda,\Lambda^{\prime}}\mathsf{R}_{\Lambda,\Lambda^{\prime}}$,
one can easily show that the so-redefined operator also satisfies the
composition law
$\mathsf{R}^{\prime}_{\Lambda,\Lambda^{\prime}}\mathsf{R}^{\prime}_{\Lambda^{\prime},\Lambda^{\prime\prime}}=\mathsf{R}^{\prime}_{\Lambda,\Lambda^{\prime\prime}}$
and the initial condition $\mathsf{R}^{\prime}_{\Lambda,\Lambda}=\id$
provided a numerical factor $a_{\Lambda,\Lambda^{\prime}}$ satisfies
$a_{\Lambda,\Lambda^{\prime}}a_{\Lambda^{\prime},\Lambda^{\prime\prime}}=a_{\Lambda,\Lambda^{\prime\prime}}$
and $a_{\Lambda,\Lambda}=1$. In the ERG analysis, it is customary to
fix this normalization ambiguity $a_{\Lambda,\Lambda^{\prime}}$ by
imposing that the ERG transformation should preserve the canonical
partition function. Under this convention, the following equality
should hold for any
$\Lambda,\Lambda^{\prime}(\leq\Lambda_{0})$:\footnote{This is just a
  convention. For partition-function-nonpreserving ERG
  transformations, eq.~\eqref{eq:65} should be replaced by
  $(1/Z_{\Lambda,\Lambda_{0}})\int_{Q}d\phi\e^{-S_{\Lambda,\Lambda_{0}}[\phi]}=(1/Z_{\Lambda^{\prime},\Lambda_{0}})\int_{Q}d\phi\e^{-S_{\Lambda^{\prime},\Lambda_{0}}[\phi]}$,
  where
  $Z_{\Lambda,\Lambda_{0}}=\int_{Q}d\phi\e^{-S_{\Lambda,\Lambda_{0}}[\phi]}$
  for any $\Lambda\leq\Lambda_{0}$. This modifies the normalization
  condition \eqref{eq:67} to
  $\int_{Q}d\phi\,\mathsf{R}_{\Lambda,\Lambda^{\prime}}[\phi,\phi^{\prime}]=Z_{\Lambda,\Lambda_{0}}/Z_{\Lambda^{\prime},\Lambda_{0}}$.
  However, the argument in \eqref{eq:69} still holds, confirming that
  the result \eqref{eq:70} is robust and independent of the
  normalization convention.}
\begin{align}
  \int_{Q}d\phi\e^{-S_{\Lambda,\Lambda_{0}}[\phi]}=\int_{Q}d\phi\e^{-S_{\Lambda^{\prime},\Lambda_{0}}[\phi]}.\label{eq:65}
\end{align}
Substituting \eqref{eq:16} on the left-hand side, we see that
\eqref{eq:65} is equivalent to
\begin{align}
  \int_{Q}d\phi^{\prime}\left(\int_{Q}d\phi\,\mathsf{R}_{\Lambda,\Lambda^{\prime}}[\phi,\phi^{\prime}]-1\right)\e^{-S_{\Lambda^{\prime},\Lambda_{0}}[\phi^{\prime}]}=0.\label{eq:66}
\end{align}
Obviously, the necessary and sufficient condition for which
\eqref{eq:66} holds for an arbitrary Boltzmann weight
$\e^{-S_{\Lambda^{\prime},\Lambda_{0}}[\phi^{\prime}]}$ is the
following condition for any $\phi^{\prime}\in Q$:
\begin{align}
  \int_{Q}d\phi\,\mathsf{R}_{\Lambda,\Lambda^{\prime}}[\phi,\phi^{\prime}]=1.\label{eq:67}
\end{align}
This is the normalization condition for the
partition-function-preserving ERG transformation. This condition gives
a stringent constraint on the representation $D$ as well as the kernel
$\widetilde{\mathsf{R}}_{\Lambda,\Lambda^{\prime}}[\ast,\ast]$.

To see this, consider the following functional:
\begin{align}
  N_{\Lambda,\Lambda^{\prime}}[\phi^{\prime}]=\int_{Q}d\phi\,\mathsf{R}_{\Lambda,\Lambda^{\prime}}[\phi,\phi^{\prime}].\label{eq:68}
\end{align}
By using the twisted boundary condition \eqref{eq:31b}, we easily find
that this functional satisfies
\begin{align}
  N_{\Lambda,\Lambda^{\prime}}[\gamma\phi^{\prime}]
  &=\int_{Q}d\phi\,\mathsf{R}_{\Lambda,\Lambda^{\prime}}[\phi,\gamma\phi^{\prime}]\nonumber\\
  &=\int_{Q}d\phi\,\mathsf{R}_{\Lambda,\Lambda^{\prime}}[\phi,\phi^{\prime}]D(\gamma^{-1})\nonumber\\
  &=N_{\Lambda,\Lambda^{\prime}}[\phi^{\prime}]D(\gamma^{-1})\label{eq:69}
\end{align}
for any $\gamma\in\Gamma$. Hence, if
$N_{\Lambda,\Lambda^{\prime}}[\phi^{\prime}]=1$ for every
$\phi^{\prime}\in Q$, we have
$N_{\Lambda,\Lambda^{\prime}}[\phi^{\prime}]=D(\gamma^{-1})$ for every
$\phi^{\prime}\in\gamma Q=\{\gamma\phi\mid\phi\in Q\}$; that is,
\eqref{eq:68} becomes discontinuous with respect to $\phi^{\prime}$ if
$D(\gamma^{-1})\neq1$. But we expect that \eqref{eq:68} should be a
continuous functional of $\phi^{\prime}$ because the ERG kernel
$\mathsf{R}_{\Lambda,\Lambda^{\prime}}[\phi,\phi^{\prime}]$ is
normally a continuous functional of $\phi^{\prime}$ (even when its
domain is extended to $\widetilde{Q}$). Hence the discontinuity of
\eqref{eq:68} contradicts the continuity of the ERG kernel. So we must
have
\begin{align}
  D(\gamma)=1\label{eq:70}
\end{align}
for any $\gamma\in\Gamma$; that is, $D$ must be the trivial
representation. Note that the trivial representation satisfies the
reflection reality \eqref{eq:60}. Note also that, when $D$ is the
trivial representation, the left-hand side of \eqref{eq:67} can be
written as
\begin{align}
  \int_{Q}d\phi\,\mathsf{R}_{\Lambda,\Lambda^{\prime}}[\phi,\phi^{\prime}]
  &=\int_{\widetilde{Q}/\Gamma}d\phi\sum_{\gamma\in\Gamma}\widetilde{\mathsf{R}}_{\Lambda,\Lambda^{\prime}}[\phi,\gamma\phi^{\prime}]\nonumber\\
  &=\int_{\widetilde{Q}/\Gamma}d\phi\sum_{\gamma\in\Gamma}\widetilde{\mathsf{R}}_{\Lambda,\Lambda^{\prime}}[\gamma^{-1}\phi,\phi^{\prime}]\nonumber\\
  &=\int_{\widetilde{Q}}d\phi\,\widetilde{\mathsf{R}}_{\Lambda,\Lambda^{\prime}}[\phi,\phi^{\prime}],\label{eq:71}
\end{align}
where the first equality follows from \eqref{eq:26}, the second
equality follows from the $\Gamma$-invariance \eqref{eq:28c}, and the
last equality follows from the identity \eqref{eq:24}. Hence, the
normalization condition \eqref{eq:67} is equivalent to the following
normalization condition for the kernel
$\widetilde{\mathsf{R}}_{\Lambda,\Lambda^{\prime}}[\ast,\ast]$:
\begin{align}
  \int_{\widetilde{Q}}d\phi\,\widetilde{\mathsf{R}}_{\Lambda,\Lambda^{\prime}}[\phi,\phi^{\prime}]=1.\label{eq:72}
\end{align}

Summarizing this section, we have seen that the ERG kernel
\eqref{eq:26} satisfies the composition law \eqref{eq:27a}, the
initial condition \eqref{eq:27b}, the reflection reality
\eqref{eq:62}, and the normalization condition \eqref{eq:67} if the
weight factor is given by the trivial representation \eqref{eq:70} and
if the kernel
$\widetilde{\mathsf{R}}_{\Lambda,\Lambda^{\prime}}[\ast,\ast]$
fulfills the composition law \eqref{eq:28a}, the initial condition
\eqref{eq:28b}, the $\Gamma$-invariance \eqref{eq:28c}, the reflection
reality \eqref{eq:63}, and the normalization condition
\eqref{eq:72}. We emphasize that the fact that $D$ is the trivial
representation is not trivial in a literal sense. As we will see in
the next section, the normalization condition \eqref{eq:70}
corresponds to a flux-quantization condition for a background magnetic
flux penetrating the multiply connected space $Q$.

\section{Wilson-loop realization of weight factors}
\label{section:4}
In quantum mechanics on $S^{1}$ discussed in section \ref{section:2},
the weight factor $D(n)=\e^{in\theta}$ is nothing but an Aharonov-Bohm
phase and can be written as a Wilson loop
$\exp(i\oint\!\!\mathscr{A})$ along a closed path winding around the
circle $n$ times, where $\mathscr{A}$ is a background Abelian gauge
field that describes a magnetic flux penetrating the circle. Likewise,
as discussed by Wu and Zee in the 1980s \cite{Wu:1984bi,Wu:1985tx}, in
several Lorentzian quantum-field-theory models whose configuration
space is multiply connected, the weight factor can also be written as
a Wilson loop with respect to a background Abelian gauge field in the
configuration space. In this section, we will discuss that this is
also the case for the Euclidean ERG: the weight factor can be written
as a Wilson loop and interpreted as an Aharonov-Bohm phase in
one-particle Euclidean quantum mechanics on the infinite-dimensional
configuration space $Q$.\footnote{A crucial difference between
  Lorentzian quantum field theory and the Euclidean ERG is that the
  $k$th homotopy group of the configuration space in the Lorentzian
  theory can correspond to the $(k-1)$st homotopy group in the
  Euclidean ERG. This arises because, in the path-integral formulation
  of Lorentzian quantum field theory, a field is defined as a map from
  a $(d-1)$-dimensional time slice $\Sigma$ of a $d$-dimensional
  spacetime $X$ to a target space $Y$. Consequently, the configuration
  space of Lorentzian quantum field theory (subject to a Dirichlet
  boundary condition at a base point) is given by the based mapping
  space $Q_{\text{Mink}}=\Map_{\ast}(\Sigma,Y)$, rather than
  $Q=\Map_{\ast}(X,Y)$. For instance, in the $d$-dimensional
  Lorentzian $O(n)$ nonlinear sigma model with an isotropic boundary
  condition at spatial infinity, the configuration space is
  $Q_{\text{Mink}}=\Map_{\ast}(S^{d-1},S^{n-1})$. Its $k$th homotopy
  group is then
  $\pi_{k}(Q_{\text{Mink}})\cong\pi_{k+d-1}(S^{n-1})\cong\pi_{k-1}(Q)$;
  see \eqref{eq:3}. Note that a nontrivial $\pi_{1}(Q_{\text{Mink}})$
  indicates the presence of a $\theta$-term.} In this picture, the
normalization condition corresponds to the flux quantization condition
$\oint\!\!\mathscr{A}\in2\pi\mathbb{Z}$, which turns out to give the
level-quantization condition for Wess-Zumino-Witten terms in nonlinear
sigma models. We will also discuss alternative gauge-equivalent
descriptions of the ERG kernel and the flow equation.

\subsection{Line integrals on based mapping spaces}
\label{section:4.1}
\begin{figure}[t]
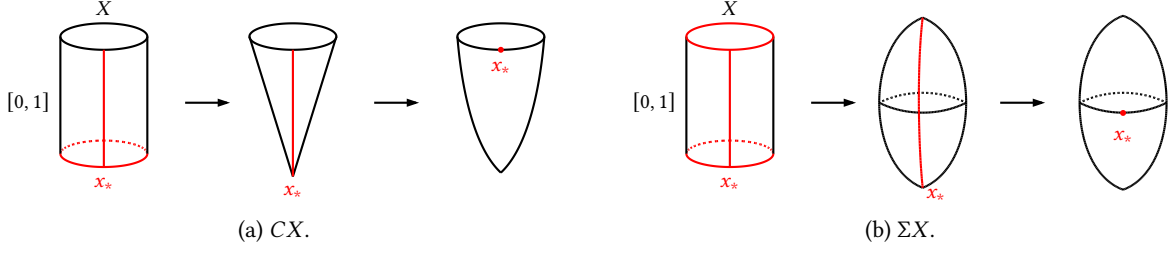

  \centering
  \begin{subfigure}[b]{0.48\textwidth}
    \centering%
    \input{figure2a.eepic}%
    \caption{$CX$.}%
    \label{figure:2a}%
  \end{subfigure}
  \hfill%
  \begin{subfigure}[b]{0.48\textwidth}
    \centering%
    \input{figure2b.eepic}%
    \caption{$\Sigma X$.}%
    \label{figure:2b}%
  \end{subfigure}
  \caption{Reduced cone $CX$ and reduced suspension $\Sigma X$ over
    the Euclidean spacetime $X$. In these diagrams, $X$ is depicted as
    a circle, making $[0,1]\times X$ a cylinder. (a) The first arrow
    indicates the collapsing of the red lower circle $\{0\}\times X$
    to the single base point $\{0\}\times\{x_{\ast}\}$, while the
    second arrow shows the collapsing of the red vertical line
    $[0,1]\times\{x_{\ast}\}$ to $\{1\}\times\{x_{\ast}\}$. This
    procedure yields $CX$, which is equivalent to the
    $(d+1)$-dimensional disk $D^{d+1}$ for $X=S^{d}$. (b) The first
    arrow represents the shrinking of both the red lower and upper
    circles ($\{0\}\times X$ and $\{1\}\times X$) to their respective
    base points. The second arrow shows the contraction of the red
    vertical line $[0,1]\times\{x_{\ast}\}$ to a single base point,
    forming $\Sigma X$. For $X=S^{d}$, $\Sigma X$ is equivalent to
    $S^{d+1}$.}
  \label{figure:2}
\end{figure}

To begin with, let us again recall the basics of paths and loops on
the based mapping space
$Q=\Map_{\ast}(X,Y)=\{\phi:X\to Y\mid\phi(x_{\ast})=y_{\ast}\}$ in
algebraic topology from a slightly different viewpoint of section
\ref{section:3.3}. (A generalization to the unbased mapping space is
straightforward.) Let $\phi_{\ast}:X\to Y$ be the
base-point-preserving constant map defined by
$\phi_{\ast}(x)=y_{\ast}$ for any $x\in X$. Let $\ell$ be an open path
on $Q$ starting from $\phi_{\ast}$ at time $t=0$ and ending at an
arbitrary $\phi$ at time $t=1$; that is, let $\ell$ be the map
\begin{align}
  \ell:[0,1]\to\Map_{\ast}(X,Y)\label{eq:73}
\end{align}
satisfying the boundary conditions $\ell(0)=\phi_{\ast}$ and
$\ell(1)=\phi$. Given such a path, we can define the map
\begin{align}
  \Phi_{\ell}:[0,1]\times X\to Y\label{eq:74}
\end{align}
by
\begin{align}
  \Phi_{\ell}(t,x)=\ell(t)(x).\label{eq:75}
\end{align}
As easily checked, this map satisfies the following boundary
conditions for any $x\in X$ and $t\in[0,1]$:
\begin{subequations}
  \begin{align}
    \Phi_{\ell}(0,x)&=\ell(0)(x)=\phi_{\ast}(x)=y_{\ast},\label{eq:76a}\\
    \Phi_{\ell}(1,x)&=\ell(1)(x)=\phi(x),\label{eq:76b}\\
    \Phi_{\ell}(t,x_{\ast})&=\ell(t)(x_{\ast})=y_{\ast},\label{eq:76c}
  \end{align}
\end{subequations}
where the last equality of \eqref{eq:76c} follows from the fact that
$\ell(t)\in\Map_{\ast}(X,Y)$ is a base-point-preserving map.
Eq.~\eqref{eq:76a} says that the map $\Phi_{\ell}$ sends the subspace
$\{0\}\times X\subset[0,1]\times X$ to a single base point
$y_{\ast}$. Likewise, eq.~\eqref{eq:76c} says that the map
$\Phi_{\ell}$ also sends the subspace
$[0,1]\times\{x_{\ast}\}\subset[0,1]\times X$ to the same point
$y_{\ast}$. Hence we can introduce an equivalence relation
$(t,x)\sim(t^{\prime},x^{\prime})$ on $[0,1]\times X$ by
$(t,x)=(t^{\prime},x^{\prime})$ or
$(t,x),(t^{\prime},x^{\prime})\in(\{0\}\times
X)\cup([0,1]\times\{x_{\ast}\})$. Correspondingly, we can define the
following induced map:
\begin{align}
  \widetilde{\Phi}_{\ell}:CX\to Y,\label{eq:77}
\end{align}
where $CX$ is the quotient space with respect to the above equivalence
relation; that is, it is defined by shrinking the subspace
$(\{0\}\times X)\cup([0,1]\times\{x_{\ast}\})$ to a single base point:
\begin{align}
  CX=\frac{[0,1]\times X}{(\{0\}\times X)\cup([0,1]\times\{x_{\ast}\})}.\label{eq:78}
\end{align}
In algebraic topology, this space is called the \textit{reduced cone}
over $X$. For instance, for $X=S^{d}$, the reduced cone is
topologically equivalent to the $(d+1)$-disk, $CS^{d}\approx D^{d+1}$;
see figure \ref{figure:2a}. In field theory, the reduced cone $CX$
provides a spacetime of one dimension higher in which
Wess-Zumino-Witten terms are defined, and the map
$\widetilde{\Phi}_{\ell}$ gives an extension of the field $\phi$ that
satisfies the boundary condition
$\widetilde{\Phi}_{\ell}|_{\partial CX}=\phi$, which is just
\eqref{eq:76b}. By construction, it is obvious that the extension
\eqref{eq:77} is essentially the open path $\ell$ on the configuration
space $Q$.

Let us next consider a closed path $\ell:[0,1]\to\Map_{\ast}(X,Y)$
satisfying the boundary conditions $\ell(0)=\ell(1)=\phi_{\ast}$. In
this case, we can also define the map $\Phi_{\ell}:[0,1]\times X\to Y$
by $\Phi_{\ell}(t,x)=\ell(t)(x)$. But the boundary conditions for
$\Phi_{\ell}$ are changed as follows:
\begin{subequations}
  \begin{align}
    \Phi_{\ell}(0,x)&=\ell(0)(x)=\phi_{\ast}(x)=y_{\ast},\label{eq:79a}\\
    \Phi_{\ell}(1,x)&=\ell(1)(x)=\phi_{\ast}(x)=y_{\ast},\label{eq:79b}\\
    \Phi_{\ell}(t,x_{\ast})&=\ell(t)(x_{\ast})=y_{\ast}.\label{eq:79c}
  \end{align}
\end{subequations}
Hence, in this case $\Phi_{\ell}$ sends the subspace
$(\{0\}\times X)\cup(\{1\}\times X)\cup([0,1]\times\{x_{\ast}\})$ into
a single base point $y_{\ast}$. Thus it induces the map
\begin{align}
  \widetilde{\Phi}_{\ell}:\Sigma X\to Y,\label{eq:80}
\end{align}
where $\Sigma X$ is defined by shrinking the subspace
$(\{0\}\times X)\cup(\{1\}\times X)\cup([0,1]\times\{x_{\ast}\})$ to a
single base point:
\begin{align}
  \Sigma X=\frac{[0,1]\times X}{(\{0\}\times X)\cup(\{1\}\times X)\cup([0,1]\times\{x_{\ast}\})}.\label{eq:81}
\end{align}
This space is called the \textit{reduced suspension} over $X$. For
$X=S^{d}$, the reduced suspension is topologically equivalent to the
$(d+1)$-sphere, $\Sigma S^{d}\approx S^{d+1}$; see figure
\ref{figure:2b}. As we will see shortly, the reduced suspension
$\Sigma X$ provides a spacetime of one dimension higher in which
Wilson-loop integrals are defined.

Now, let us next consider line integrals along a path $\ell$ on the
configuration space $Q$ by following (and refining) the argument of Wu
and Zee \cite{Wu:1984bi,Wu:1985tx}. To this end, we first introduce
the following functional-differential one-form on $Q$:
\begin{align}
  \mathscr{A}=\int_{X}dx\,\mathscr{A}(\phi(x))\delta\phi(x),\label{eq:82}
\end{align}
where $\mathscr{A}(\phi(x))$ is an arbitrary function of the field
$\phi(x)$ and its $x$-derivatives. We wish to define a line integral
of \eqref{eq:82} along a path $\ell$. As in the case of ordinary
differential forms, this can be done by the pullback of the
functional-differential one-form $\mathscr{A}$ by the path $\ell$. For
the component $\mathscr{A}(\phi(x))$, the pullback means to replace
$\phi(x)$ by $\ell(t)(x)$. Thus we have
$\mathscr{A}(\phi(x))\to\mathscr{A}(\ell(t)(x))=\mathscr{A}(\Phi_{\ell}(t,x))$,
where we have used \eqref{eq:75}. For the basis
functional-differential one-form $\delta\phi(x)$, on the other hand,
the pullback means to evaluate the variation $\delta\phi(x)$ along the
path $\ell$. Thus we have
$\delta\phi(x)\to\ell(t+dt)(x)-\ell(t)(x)=\Phi_{\ell}(t+dt,x)-\Phi_{\ell}(t,x)=(\partial_{t}\Phi_{\ell})(t,x)dt$. Putting
these things together, we arrive at the following differential
one-form on the interval $[0,1]$:
\begin{align}
  \ell^{\,\ast}\mathscr{A}=\left(\int_{X}dx\,\mathscr{A}(\Phi_{\ell}(t,x))(\partial_{t}\Phi_{\ell})(t,x)\right)dt,\label{eq:83}
\end{align}
where $\ell^{\,\ast}\mathscr{A}$ denotes the pullback of $\mathscr{A}$
by $\ell$. The line integral of the functional-differential one-form
$\mathscr{A}$ along the path $\ell$ is now defined by
\begin{align}
  \int_{\ell}\mathscr{A}
  =\int_{[0,1]}\ell^{\,\ast}\mathscr{A}
  =\int_{[0,1]\times X}dtdx\,\mathscr{A}(\Phi_{\ell}(t,x))(\partial_{t}\Phi_{\ell})(t,x).\label{eq:84}
\end{align}
Note that this definition satisfies the following desired properties
of line integrals:
\begin{subequations}
  \begin{align}
    \int_{\ell_{1}\ast\ell_{2}}\mathscr{A}&=\int_{\ell_{1}}\mathscr{A}+\int_{\ell_{2}}\mathscr{A},\label{eq:85a}\\
    \int_{\ell^{-1}}\mathscr{A}&=-\int_{\ell}\mathscr{A},\label{eq:85b}
  \end{align}
\end{subequations}
which follow from
$\Phi_{\ell_{1}\ast\ell_{2}}(t,x)=\Phi_{\ell_{1}}(2t,x)$ for
$0\leq t\leq1/2$,
$\Phi_{\ell_{1}\ast\ell_{2}}(t,x)=\Phi_{\ell_{2}}(2t-1,x)$ for
$1/2\leq t\leq1$, and $\Phi_{\ell^{-1}}(t,x)=\Phi_{\ell}(1-t,x)$ for
any $0\leq t\leq1$. Note also that, if the component
$\mathscr{A}(\phi(x))$ is a pseudo-tensor, it changes its sign under
the reflection. In this case, the line integral along the reflected
path $\vartheta\ell$ satisfies the following identity:
\begin{align}
  \int_{\vartheta\ell}\mathscr{A}=-\int_{\ell}\mathscr{A}\quad\text{if $\mathscr{A}$ is a pseudo-tensor}.\label{eq:86}
\end{align}

Now, the last integral of \eqref{eq:84} is the integral over the
cylinder $[0,1]\times X$. But it can be reduced to the integral over
the reduced cone $CX$ (reduced suspension $\Sigma X$) for an open (a
closed) path $\ell$ if $\mathscr{A}(\Phi_{\ell})$ is proportional to
the $x$-derivative $\partial_{x}\Phi_{\ell}$. For instance, for an
open path $\ell$ we have the boundary conditions
$\Phi_{\ell}(0,x)=y_{\ast}$ and $\Phi_{\ell}(t,x_{\ast})=y_{\ast}$ for
any $x$ and $t$, from which we find $(\partial_{x}\Phi_{\ell})(0,x)=0$
and $(\partial_{t}\Phi_{\ell})(t,x_{\ast})=0$. Hence if
$\mathscr{A}(\Phi_{\ell})\propto\partial_{x}\Phi_{\ell}$, the
integrand of \eqref{eq:84} vanishes on the subspace
$(\{0\}\times X)\cup([0,1]\times\{x_{\ast}\})$ such that the
integration domain $[0,1]\times X$ can be reduced to the reduced cone
$CX$. Likewise, for a closed path $\ell$ the integral \eqref{eq:84}
has no contribution from the subspace
$(\{0\}\times X)\cup(\{1\}\times X)\cup([0,1]\times\{x_{\ast}\})$ such
that the integration domain can be reduced to the reduced suspension
$\Sigma X$. Thus, for $\mathscr{A}(\phi(x))$ being proportional to
$(\partial_{x}\phi)(x)$, we find
\begin{align}
  \int_{\ell}\mathscr{A}=
  \begin{cases}
    \displaystyle\int_{CX}dtdx\,\mathscr{A}(\widetilde{\Phi}_{\ell}(t,x))(\partial_{t}\widetilde{\Phi}_{\ell})(t,x)&\text{for $\ell$ open},\\[1em]
    \displaystyle\int_{\Sigma X}dtdx\,\mathscr{A}(\widetilde{\Phi}_{\ell}(t,x))(\partial_{t}\widetilde{\Phi}_{\ell})(t,x)&\text{for $\ell$ closed},\\
  \end{cases}\label{eq:87}
\end{align}
where $\widetilde{\Phi}_{\ell}$ is the induced map given by
\eqref{eq:77} or \eqref{eq:80}. As we will see soon, the first line of
the right-hand side of \eqref{eq:87} gives Wess-Zumino-Witten terms in
nonlinear sigma models.

For an arbitrary functional-differential one-form $\mathscr{A}$, the
line integral \eqref{eq:87} depends on the path $\ell$ in
general. However, as in the case of ordinary differential forms, the
line integral becomes invariant under the continuous deformation of
$\ell$ if $\mathscr{A}$ is a closed form. Here the
functional-differential one-form $\mathscr{A}$ is said to be closed if
the following variation vanishes:
\begin{align}
  \delta\mathscr{A}
  &=\int_{X}dx\int_{X}dy\frac{\delta\mathscr{A}(\phi(x))}{\delta\phi(y)}\delta\phi(y)\wedge\delta\phi(x)\nonumber\\
  &=\frac{1}{2!}\int_{X}dx\int_{X}dy\left(\frac{\delta\mathscr{A}(\phi(y))}{\delta\phi(x)}-\frac{\delta\mathscr{A}(\phi(x))}{\delta\phi(y)}\right)\delta\phi(x)\wedge\delta\phi(y).\label{eq:88}
\end{align}
If $\delta\mathscr{A}=0$, the loop integral $\oint_{\ell}\mathscr{A}$
depends only on the homology class of the closed path $\ell$. In this
case the one-dimensional representation of the fundamental group
$\pi_{1}(Q)\cong\Gamma$ can be written as the following Wilson loop
(or holonomy):\footnote{We here assume that the first homology group
  $H_{1}(Q;\mathbb{Z})$ (i.e., the Abelianization of $\pi_{1}(Q)$) has
  no torsion. Since the weight factor $D$ is a one-dimensional
  representation, it is strictly Abelian and thus captures only the
  information of $H_{1}(Q;\mathbb{Z})$, rather than the full
  (non-Abelian) structure of $\pi_{1}(Q)$ \cite{Horvathy:1988vh}. It
  should be noted that any torsion part of $H_{1}(Q;\mathbb{Z})$
  cannot be realized as a Wilson loop of a gauge field. However, this
  technical restriction does not affect our physical conclusions: as
  shown in section \ref{section:3.4}, the normalization condition
  strictly requires $D(\gamma)=1$ (the trivial representation) for all
  $\gamma\in\Gamma$. Therefore, even if $H_{1}(Q;\mathbb{Z})$
  contained a torsion part like $\mathbb{Z}_{2}$, its corresponding
  weight factor would necessarily be unity, making it safely
  negligible in our present framework.}
\begin{align}
  D(\gamma_{[\ell]})=\exp\left(i\oint_{\ell}\mathscr{A}\right),\label{eq:89}
\end{align}
which satisfies the desired composition law
$D(\gamma_{[\ell_{1}]})D(\gamma_{[\ell_{2}]})=D(\gamma_{[\ell_{1}\ast\ell_{2}]})$
for any closed paths $\ell_{1},\ell_{2}$ based at $\phi_{\ast}$. In
addition, if $\mathscr{A}$ is a pseudo-tensor, the Wilson loop
satisfies the unitarity condition
$\overline{D(\gamma_{[\ell]})}=D(\gamma_{[\ell]}^{-1})$. In this case,
the normalization condition $D(\gamma_{[\ell]})=1$ corresponds to the
flux-quantization condition $\oint_{\ell}\mathscr{A}\in2\pi\mathbb{Z}$
for the background magnetic flux described by the
functional-differential one-form \eqref{eq:82}.

Let us next present a simple explicit example of the above formulation
by using nonlinear sigma models.

\subsection{Example: \texorpdfstring{$d$}{d}-dimensional
  \texorpdfstring{$O(d+2)$}{O(d+2)} nonlinear sigma model}
\label{section:4.2}
Let us consider the $d$-dimensional $O(d+2)$ nonlinear sigma model on
the Euclidean spacetime $\mathbb{R}^{d}$. In this model, the field is
a map $\phi=(\phi^{1},\cdots,\phi^{d+2}):\mathbb{R}^{d}\to S^{d+1}$
satisfying
\begin{align}
  (\phi^{1}(x))^{2}+\cdots+(\phi^{d+2}(x))^{2}=R^{2}\label{eq:90}
\end{align}
for any $x\in\mathbb{R}^{d}$, where $R>0$ is the radius of $S^{d+1}$
and has mass dimension $(d-2)/2$. We impose the following isotropic
boundary condition at infinity:
\begin{align}
  \lim_{|x|\to\infty}\phi(x)=y_{\ast},\label{eq:91}
\end{align}
where $y_{\ast}=(y_{\ast}^{1},\cdots,y_{\ast}^{d+2})\in S^{d+1}$ is an
arbitrary constant. As already discussed in section \ref{section:1},
under this boundary condition the domain of $\phi$ contains the
infinity such that the spacetime for $\phi$ effectively becomes the
one-point compactification of $\mathbb{R}^{d}$, which is topologically
equivalent to the $d$-sphere,
$\mathbb{R}^{d}\cup\{\infty\}\approx S^{d}$. Hence, the configuration
space of the $d$-dimensional $O(d+2)$ nonlinear sigma model is the
based mapping space
$Q=\Map_{\ast}(S^{d},S^{d+1})=\{\phi:S^{d}\to
S^{d+1}\mid\phi(\infty)=y_{\ast}\}$, whose fundamental group is given
by (see \eqref{eq:3})
\begin{align}
  \pi_{1}(Q)\cong\pi_{d+1}(S^{d+1})\cong\mathbb{Z}.\label{eq:92}
\end{align}
Hence $Q$ is multiply connected. Below we will show that the Wilson
loop \eqref{eq:89} is given by the Wess-Zumino-Witten term.

To this end, let $\omega$ be the volume form of the target space
$S^{d+1}$ given by
\begin{align}
  \omega=\frac{1}{(d+1)!}\frac{1}{R\vol(S^{d+1})}\varepsilon^{a_{1}a_{2}\cdots a_{d+2}}\phi^{a_{1}}d\phi^{a_{2}}\wedge\cdots\wedge d\phi^{a_{d+2}},\label{eq:93}
\end{align}
where $\vol(S^{d+1})=2\pi^{(d+2)/2}R^{d+1}/\Gamma((d+2)/2)$ stands for
the volume of $S^{d+1}$. $\varepsilon^{a_{1}\cdots a_{d+2}}$ is the
totally antisymmetric tensor satisfying
$\varepsilon^{1\cdots(d+2)}=+1$. The volume form $\omega$ is
normalized so as to satisfy $\int_{S^{d+1}}\omega=1$. Then, the
Wess-Zumino-Witten term of the $d$-dimensional $O(d+2)$ nonlinear
sigma model is given by the integral of the pullback
$\widetilde{\Phi}_{\ell}^{\ast}\omega$ over the reduced cone
$CS^{d}\approx D^{d+1}$:
\begin{align}
  S_{\text{WZW}}[\ell]
  &=-i2\pi k\int_{CS^{d}}\widetilde{\Phi}_{\ell}^{\ast}\omega\nonumber\\
  &=-\frac{i2\pi k}{(d+1)!R\vol(S^{d+1})}\int_{CS^{d}}dtdx\,\varepsilon^{a_{1}\cdots a_{d+2}}\varepsilon^{\mu_{1}\cdots\mu_{d+1}}\widetilde{\Phi}_{\ell}^{a_{1}}(\partial_{\mu_{1}}\widetilde{\Phi}_{\ell}^{a_{2}})\cdots(\partial_{\mu_{d+1}}\widetilde{\Phi}_{\ell}^{a_{d+2}}),\label{eq:94}
\end{align}
where $k$ is the Wess-Zumino-Witten level and
$\widetilde{\Phi}_{\ell}=(\widetilde{\Phi}_{\ell}^{1},\cdots,\widetilde{\Phi}_{\ell}^{d+2}):CS^{d}\approx
D^{d+1}\to S^{d+1}$ is the induced map associated with an open path
$\ell$ on $Q$. An important observation here is that the
Wess-Zumino-Witten term \eqref{eq:94} can be rewritten as the
following line integral along the path $\ell$:
\begin{align}
  S_{\text{WZW}}[\ell]
  =-i\int_{\ell}\mathscr{A}
  =-i\int_{[0,1]\times S^{d}}dtdx\,\mathscr{A}^{a}(\Phi_{\ell})\partial_{t}\Phi_{\ell}^{a}.\label{eq:95}
\end{align}
Here $\mathscr{A}=\int_{S^{d}}dx\,\mathscr{A}^{a}(\phi)\delta\phi^{a}$
is an Abelian functional-differential one-form whose components are
given by the pseudo-scalars
\begin{align}
  \mathscr{A}^{a}(\phi)=\frac{2\pi k}{d!R\vol(S^{d+1})}\varepsilon^{a_{1}\cdots a_{d+1}a}\varepsilon^{\mu_{1}\cdots\mu_{d}}(\partial_{\mu_{1}}\phi^{a_{1}})\cdots(\partial_{\mu_{d}}\phi^{a_{d}})\phi^{a_{d+1}},\label{eq:96}
\end{align}
and $\Phi_{\ell}:[0,1]\times S^{d}\to S^{d+1}$ is the map defined by
\eqref{eq:75}. Notice that the components \eqref{eq:96} vanish on the
subspace $(\{0\}\times S^{d})\cup([0,1]\times\{\infty\})$. So the
integral over the cylinder $[0,1]\times S^{d}$ reduces to the integral
over the reduced cone $CS^{d}\approx D^{d+1}$. Note also that the
functional-differential one-form $\mathscr{A}$ is closed. Indeed, a
straightforward calculation gives
\begin{align}
  \delta\mathscr{A}
  &=\frac{1}{2!}\int_{S^{d}}dx\int_{S^{d}}dy\left(\frac{\delta\mathscr{A}^{b}(\phi(y))}{\delta\phi^{a}(x)}-\frac{\delta\mathscr{A}^{a}(\phi(x))}{\delta\phi^{b}(y)}\right)\delta\phi^{a}(x)\wedge\delta\phi^{b}(y)\nonumber\\
  &\propto\int_{S^{d}}dx\,\varepsilon^{\mu_{1}\cdots\mu_{d}}\varepsilon^{a_{1}\cdots a_{d}ab}(\partial_{\mu_{1}}\phi^{a_{1}})(x)\cdots(\partial_{\mu_{d}}\phi^{a_{d}})(x)\delta\phi^{a}(x)\wedge\delta\phi^{b}(x)\nonumber\\
  &=0,\label{eq:97}
\end{align}
where we have used the fact that the quantity
$\varepsilon^{a_{1}\cdots
  a_{d}ab}(\partial_{\mu_{1}}\phi^{a_{1}})\cdots(\partial_{\mu_{d}}\phi^{a_{d}})\delta\phi^{a}\wedge\delta\phi^{b}$
identically vanishes because it totally antisymmetrizes $d+2$ tangent
vectors ($\partial_{\mu}\phi$ and $\delta\phi$) within the
$(d+1)$-dimensional tangent space of $S^{d+1}$.\footnote{Both
  $\partial_{\mu}\phi=(\partial_{\mu}\phi^{1},\cdots,\partial_{\mu}\phi^{d+2})$
  and $\delta\phi=(\delta\phi^{1},\cdots,\delta\phi^{d+2})$ are
  orthogonal to $\phi=(\phi^{1},\cdots,\phi^{d+2})\in S^{d+1}$ and
  satisfy $\phi^{a}\partial_{\mu}\phi^{a}=0$ and
  $\phi^{a}\delta\phi^{a}=0$. Any $d+2$ vectors are linearly dependent
  in the $(d+1)$-dimensional vector space.} Hence the line integral
\eqref{eq:95} just depends on the homotopy class of the path
$\ell$. Notice that, though the field strength
$\mathscr{F}=\delta\mathscr{A}$ vanishes on $Q$, the line integral
along a closed path $\ell$ is nonvanishing in general. Indeed, by
using the degree formula (see, e.g., Section 6.2 of
\cite{Flanders:1989}), we find
\begin{align}
  \oint_{\ell}\mathscr{A}
  &=\int_{\Sigma S^{d}}dtdx\,\mathscr{A}(\widetilde{\Phi}_{\ell}(t,x))(\partial_{t}\widetilde{\Phi}_{\ell})(t,x)\nonumber\\
  &=2\pi k\int_{\Sigma S^{d}}\widetilde{\Phi}_{\ell}^{\ast}\omega\nonumber\\
  &=2\pi k\deg(\widetilde{\Phi}_{\ell})\int_{S^{d+1}}\omega\nonumber\\
  &=2\pi k\deg(\widetilde{\Phi}_{\ell}),\label{eq:98}
\end{align}
where $\deg(\widetilde{\Phi}_{\ell})\in\mathbb{Z}$ stands for the
degree of the induced map
$\widetilde{\Phi}_{\ell}:\Sigma S^{d}\approx S^{d+1}\to S^{d+1}$ that
describes the winding number of the closed path $\ell$. Since the
degree is an integer, the flux-quantization condition
$\oint_{\ell}\mathscr{A}\in2\pi\mathbb{Z}$ requires the parameter $k$
to be an integer. This is the level quantization for the
Wess-Zumino-Witten term from the Wilson-loop perspective. Notice that,
irrespective of whether $k$ is quantized or not, the level $k$ can
never be renormalized under the ERG.

\subsection{Alternative gauge-equivalent forms of ERG kernel and flow
  equation}
\label{section:4.3}
One advantage of the above Wilson-loop perspective is that we can
remove the Wess-Zumino-Witten term from the Wilson action by an
Abelian gauge transformation and rewrite the theory in an alternative
gauge-equivalent form. Suppose that the configuration space $Q$ is
multiply connected and the Wilson action
$S_{\Lambda,\Lambda_{0}}[\phi]$ contains the Wess-Zumino-Witten term
$-i\int_{\phi_{\ast}}^{\phi}\mathscr{A}$, where
$\int_{\phi_{\ast}}^{\phi}$ stands for a line integral along a path
from $\phi_{\ast}$ to $\phi$. Then we define a new Wilson action
$S^{\prime}_{\Lambda,\Lambda_{0}}[\phi]$ by
\begin{align}
  \e^{-S^{\prime}_{\Lambda,\Lambda_{0}}[\phi]}=\exp\left(-i\int_{\phi_{\ast}}^{\phi}\mathscr{A}\right)\e^{-S_{\Lambda,\Lambda_{0}}[\phi]}.\label{eq:99}
\end{align}
By definition, this new Wilson action does not contain the
Wess-Zumino-Witten term anymore. But the ERG kernel becomes dependent
on the Wess-Zumino-Witten term even if the weight factor is the
trivial representation. Indeed, the ERG kernel for the new Boltzmann
weight \eqref{eq:99} is given by
\begin{align}
  \mathsf{R}^{\prime}_{\Lambda,\Lambda^{\prime}}[\phi,\phi^{\prime}]
  &=\exp\left(-i\int_{\phi_{\ast}}^{\phi}\mathscr{A}\right)\mathsf{R}_{\Lambda,\Lambda^{\prime}}[\phi,\phi^{\prime}]\exp\left(i\int_{\phi_{\ast}}^{\phi^{\prime}}\mathscr{A}\right)\nonumber\\
  &=\sum_{\gamma\in\Gamma}\exp\left(-i\int_{\phi_{\ast}}^{\phi}\mathscr{A}\right)\exp\left(i\int_{\phi_{\ast}}^{\gamma\phi_{\ast}}\mathscr{A}\right)\exp\left(i\int_{\phi_{\ast}}^{\phi^{\prime}}\mathscr{A}\right)\widetilde{\mathsf{R}}_{\Lambda,\Lambda^{\prime}}[\phi,\gamma\phi^{\prime}]\nonumber\\
  &=\sum_{\gamma\in\Gamma}\exp\left(-i\int_{\gamma\phi^{\prime}}^{\phi}\mathscr{A}\right)\widetilde{\mathsf{R}}_{\Lambda,\Lambda^{\prime}}[\phi,\gamma\phi^{\prime}],\label{eq:100}
\end{align}
where the second equality follows from \eqref{eq:26} and the
Wilson-loop realization of the weight factor
$D(\gamma)=\exp(i\int_{\phi_{\ast}}^{\gamma\phi_{\ast}}\mathscr{A})$.
We note that, irrespective of whether the flux is quantized or not,
this new kernel satisfies the following periodic boundary conditions:
\begin{subequations}
  \begin{align}
    \mathsf{R}^{\prime}_{\Lambda,\Lambda^{\prime}}[\gamma\phi,\phi^{\prime}]&=\mathsf{R}^{\prime}_{\Lambda,\Lambda^{\prime}}[\phi,\phi^{\prime}],\label{eq:101a}\\
    \mathsf{R}^{\prime}_{\Lambda,\Lambda^{\prime}}[\phi,\gamma\phi^{\prime}]&=\mathsf{R}^{\prime}_{\Lambda,\Lambda^{\prime}}[\phi,\phi^{\prime}].\label{eq:101b}
  \end{align}
\end{subequations}
Note also that the ERG generator for the new Boltzmann weight is given
by the following similarity transformation (Abelian gauge
transformation):
\begin{align}
  \mathsf{G}^{\prime}_{\Lambda}[\phi,\tfrac{\delta}{\delta\phi}]
  &=\exp\left(-i\int_{\phi_{\ast}}^{\phi}\mathscr{A}\right)\mathsf{G}_{\Lambda}[\phi,\tfrac{\delta}{\delta\phi}]\exp\left(+i\int_{\phi_{\ast}}^{\phi}\mathscr{A}\right)\nonumber\\
  &=\mathsf{G}_{\Lambda}[\phi,\tfrac{\delta}{\delta\phi}+i\mathscr{A}].\label{eq:102}
\end{align}
The ERG flow equation for the new Wilson action is therefore
\begin{align}
  -\Lambda\frac{\partial}{\partial\Lambda}\e^{-S^{\prime}_{\Lambda,\Lambda_{0}}[\phi]}=\mathsf{G}_{\Lambda}[\phi,\tfrac{\delta}{\delta\phi}+i\mathscr{A}]\e^{-S^{\prime}_{\Lambda,\Lambda_{0}}[\phi]}.\label{eq:103}
\end{align}
Hence there are two equivalent descriptions for the ERG transformation
in Euclidean field theory whose configuration space is multiply
connected. One is the ERG transformation given by the kernel
\eqref{eq:26} (with $D(\gamma)=1$) that acts on the Wilson action
containing a Wess-Zumino-Witten term, and the other is the ERG
transformation given by the kernel \eqref{eq:100} that acts on the
Wilson action with no Wess-Zumino-Witten term. These two descriptions
are related through the Abelian gauge transformation in the
infinite-dimensional configuration space and hence physically
equivalent.

To summarize, we have seen that the weight factor can be written as a
Wilson loop \eqref{eq:89} of a background Abelian gauge field on the
configuration space. The normalization condition \eqref{eq:70} gives
the flux-quantization condition, resulting in the level quantization
of Wess-Zumino-Witten terms. Since the weight factor remains unchanged
under the ERG transformation, the level is exactly nonrenormalized
even if it is not quantized.

\section{Concluding remarks}
\label{section:5}
In the literature, ERG transformations are studied mostly in their
infinitesimal forms---the flow equations. However, the flow equations
are local functional-differential equations on the configuration space
$Q$; therefore, they cannot describe the global structure of $Q$
unless we specify boundary conditions correctly. In this paper, we
have developed a general theory of the finite ERG transformation for
Euclidean field theory whose configuration space is multiply
connected. We have first introduced the equivalence relation on the
set of Boltzmann weights and then shown that the ERG kernel on a
multiply connected configuration space $Q$ is given by the linear
combination of the ERG kernels on the universal covering space
$\widetilde{Q}$. To fulfill the inhomogeneous semigroup properties,
the reflection reality, and the normalization condition, the ERG
kernel on the universal covering space must satisfy the conditions
\eqref{eq:28a}--\eqref{eq:28c}, \eqref{eq:63}, and \eqref{eq:72}. In
particular, the weight factor must be the trivial representation. We
have then discussed the Wilson-loop realization of the weight factor
as well as the gauge-equivalent alternatives of the ERG kernel and the
flow equation. Specifically, we have shown that the trivial
representation corresponds to the level quantization of
Wess-Zumino-Witten terms. Note that our general theory does not depend
on any specific formulation of the ERG; it is just based on the
assumption that the ERG transformation can be realized as the
functional-integral transform \eqref{eq:16} for any $Q$.

We note that our covering-space method could in principle be applied
to gauge theory as well. Consider, for example, the $d$-dimensional
pure Yang-Mills gauge theory whose gauge group is a Lie group $G$. The
configuration space $Q$ of this theory is the set of
gauge-inequivalent gauge fields; that is, it is given by the orbit
space
\begin{align}
  Q=\mathcal{A}/\mathcal{G},\label{eq:104}
\end{align}
where $\mathcal{A}$ is the space of gauge fields and $\mathcal{G}$ is
the group of local gauge transformations. Note that $\mathcal{A}$ is a
contractible space, meaning that all the homotopy groups of
$\mathcal{A}$ are trivial. Hence, the long exact sequence of the
homotopy groups
\begin{align}
  \cdots
  \to\pi_{k}(\mathcal{A})
  \to\pi_{k}(\mathcal{A}/\mathcal{G})
  \to\pi_{k-1}(\mathcal{G})
  \to\pi_{k-1}(\mathcal{A})
  \to\cdots\label{eq:105}
\end{align}
reduces to the exact sequence
$0\to\pi_{k}(\mathcal{A}/\mathcal{G})\to\pi_{k-1}(\mathcal{G})\to0$,
meaning that we have the isomorphism
$\pi_{k}(\mathcal{A}/\mathcal{G})\cong\pi_{k-1}(\mathcal{G})$. Suppose
that all the local gauge transformations converge to the identity
element $e\in G$ at spacetime infinity. In this case, $\mathcal{G}$ is
topologically equivalent to the based mapping space
$\Map_{\ast}(S^{d},G)=\{g:S^{d}\to G\mid g(\infty)=e\}$. Thus we have
\begin{align}
  \pi_{k}(Q)\cong\pi_{k-1}(\mathcal{G})\cong\pi_{k-1}(\Map_{\ast}(S^{d},G))\cong\pi_{d+k-1}(G).\label{eq:106}
\end{align}
In particular, $\pi_{1}(Q)\cong\pi_{0}(\mathcal{G})\cong\pi_{d}(G)$,
from which we see that the configuration space of the $d$-dimensional
pure Yang-Mills gauge theory is multiply connected if the $d$th
homotopy group of the gauge group $G$ is nontrivial. A typical example
of such a case is the three-dimensional $SU(N)$ Yang-Mills gauge
theory, where
$\pi_{1}(Q)\cong\pi_{0}(\mathcal{G})\cong\pi_{3}(SU(N))\cong\mathbb{Z}$
for any $N\geq2$. In this case, the weight factor is given by the
Chern-Simons term, and the normalization condition leads to the
quantization of the Chern-Simons level. It then follows from the
nonrenormalization theorem \eqref{eq:42} that the Chern-Simons level
is never renormalized under the ERG (even if it is not quantized). In
this way, our covering-space method and nonrenormalization theorem
could be applied to gauge theory as well. However, the construction of
the ERG kernel itself becomes a highly nontrivial problem in gauge
theory. This is due to the absence of a global gauge-fixing
\cite{Singer:1978dk} (i.e., the Gribov problem \cite{Gribov:1977wm})
even on the universal covering space $\widetilde{Q}$,\footnote{The
  absence of a global gauge-fixing on $\widetilde{Q}$ can be readily
  understood following Singer's argument \cite{Singer:1978dk}. First,
  note that $\mathcal{G}$ is not path-connected because
  $\pi_{0}(\mathcal{G})$ is nontrivial. Let $\mathcal{G}_{0}$ denote
  the path-connected component containing the identity
  element. Physically, $\mathcal{G}_{0}$ represents the group of small
  gauge transformations, whereas the other path-connected components
  correspond to large gauge transformations. The universal covering
  space of $Q$ is then given by the orbit space
  $\widetilde{Q}=\mathcal{A}/\mathcal{G}_{0}$. In the fiber-bundle
  language, the absence of a global gauge-fixing on $\widetilde{Q}$ is
  equivalent to the absence of a global section for the principal
  bundle
  $\mathcal{G}_{0}\to\mathcal{A}\to\mathcal{A}/\mathcal{G}_{0}$. This
  is equivalent to stating that this principal bundle is not the
  trivial bundle; that is, $\mathcal{A}$ cannot be written as
  $\mathcal{A}\approx(\mathcal{A}/\mathcal{G}_{0})\times\mathcal{G}_{0}$. Indeed,
  if such a decomposition were possible, we would have
  $\pi_{k}(\mathcal{A})\cong\pi_{k}(\mathcal{A}/\mathcal{G}_{0})\oplus\pi_{k}(\mathcal{G}_{0})$
  for any $k$. However, this isomorphism never holds because
  $\pi_{k}(\mathcal{A})$ is trivial for all $k$, whereas
  $\pi_{k}(\mathcal{G}_{0})\cong\pi_{k}(\mathcal{G})\cong\pi_{k+d}(G)$
  ($k\geq1$) is not. Consequently, there is no globally well-defined
  gauge-fixing function on $\widetilde{Q}$, meaning that functional
  integrals on $\widetilde{Q}$ cannot be defined via the Faddeev-Popov
  procedure with a single gauge-fixing function. This constitutes the
  topological obstruction to defining the functional integral on
  $\widetilde{Q}$.} causing a serious unsolved problem in defining the
functional integral on $\widetilde{Q}$. A second-best solution for
this problem may be to enlarge the integration domain from
$\widetilde{Q}$ to the Gribov region and implement the idea of the
refined Gribov-Zwanziger action \cite{Vandersickel:2012tz} in the ERG.

\begin{appendices}
  \setcounter{equation}{0}
  \renewcommand{\theequation}{\thesection.\arabic{equation}}
  \renewcommand{\theHequation}{\thesection.\arabic{equation}}
  \section{Simple examples}
  \label{appendix:A}
  In general, it is hard to construct the ERG kernels for generic
  configuration spaces. However, the situation gets simplified if the
  target space is a flat Euclidean space, for which we can easily
  introduce the Gaussian measure (i.e., we can easily define the free
  field theory). In this section, we will present simple examples of
  the ERG kernels that satisfy the composition law, the initial
  condition, the reflection reality, and the normalization condition
  both for simply and multiply connected configuration spaces.

  To begin with, let us consider Euclidean scalar field theory on the
  $d$-dimensional Euclidean space $\mathbb{R}^{d}$. Suppose that the
  target space is an $n$-dimensional compact flat Euclidean
  manifold. As is well known, such an $n$-manifold can be written as
  an orbit space $\mathbb{R}^{n}/\Gamma$, where
  $\Gamma\subset\Isom(\mathbb{R}^{n})$ is a discrete subgroup of the
  isometry group
  $\Isom(\mathbb{R}^{n})\cong\mathbb{R}^{n}\rtimes O(n)$ whose action
  on $\mathbb{R}^{n}$ has no fixed point; see, e.g., Section 5 of
  Chapter I\hspace{-.1em}I in \cite{Charlap:1986}.\footnote{The
    isometry group of $\mathbb{R}^{n}$ is given by the semidirect
    product of the translation group $\mathbb{R}^{n}$ and the
    orthogonal group $O(n)$. Any element of
    $\Isom(\mathbb{R}^{n})\cong\mathbb{R}^{n}\rtimes O(n)$ is
    specified by a pair $(a,R)$, where $a\in\mathbb{R}^{n}$ and
    $R\in O(n)$. It acts on $x\in\mathbb{R}^{n}$ as $(a,R)x=a+Rx$. The
    multiplication rule is
    $(a,R)(a^{\prime},R^{\prime})=(a+Ra^{\prime},RR^{\prime})$. The
    identity element is $(0,1)$ and the inverse of $(a,R)$ is
    $(a,R)^{-1}=(-R^{-1}a,R^{-1})$, where $1$ stands for the identity
    matrix.} The configuration space of such a Euclidean field theory
  with no particular boundary condition at spacetime infinity is given
  by the following (unbased) mapping space:
  \begin{align}
    Q=\Map(\mathbb{R}^{d},\mathbb{R}^{n}/\Gamma)=\{\phi:\mathbb{R}^{d}\to\mathbb{R}^{n}/\Gamma\}.\label{eq:A.1}
  \end{align}
  Notice that, since $\mathbb{R}^{d}$ is contractible, the mapping
  space $\Map(\mathbb{R}^{d},\mathbb{R}^{n}/\Gamma)$ is homotopy
  equivalent to the target space $\mathbb{R}^{n}/\Gamma$. Hence the
  $k$th homotopy group of $Q$ is isomorphic to the $k$th homotopy
  group of $\mathbb{R}^{n}/\Gamma$, which is simply given by
  \begin{align}
    \pi_{k}(Q)\cong\pi_{k}(\mathbb{R}^{n}/\Gamma)\cong
    \begin{cases}
      0&\text{for $k=0$},\\
      \Gamma&\text{for $k=1$},\\
      0&\text{for $k\geq2$}.\\
    \end{cases}\label{eq:A.2}
  \end{align}
  In addition, one can show that (i) there exists a covering map
  $p:\Map(\mathbb{R}^{d},\mathbb{R}^{n})\to\Map(\mathbb{R}^{d},\mathbb{R}^{n}/\Gamma)$
  and (ii) the mapping space $\Map(\mathbb{R}^{d},\mathbb{R}^{n})$ is
  simply connected. Hence $\Map(\mathbb{R}^{d},\mathbb{R}^{n})$ is the
  universal covering space $\widetilde{Q}$ of $Q$. Thus we find
  \begin{align}
    \widetilde{Q}=\Map(\mathbb{R}^{d},\mathbb{R}^{n}).\label{eq:A.3}
  \end{align}
  
  In the following, we will first construct the ERG kernel on
  $\widetilde{Q}$ by the standard high-momentum-mode-integration
  procedure and then construct the ERG kernel on
  $Q\approx\widetilde{Q}/\Gamma$ by the covering-space method.

  \subsection{ERG kernel on
    \texorpdfstring{$\widetilde{Q}=\Map(\mathbb{R}^{d},\mathbb{R}^{n})$}{Map(Rd,Rn)}}
  \label{appendix:A.1}
  Let $\phi=(\phi^{1},\cdots,\phi^{n})\in\widetilde{Q}$ be an
  $n$-component real scalar field on $\mathbb{R}^{d}$. The Wilson
  action $S_{\Lambda,\Lambda_{0}}[\phi]$ is assumed to be of the form
  \begin{align}
    S_{\Lambda,\Lambda_{0}}[\phi]=\frac{1}{2}\langle\phi,C_{\Lambda}^{-1}\phi\rangle+V_{\Lambda,\Lambda_{0}}[\phi],\label{eq:A.4}
  \end{align}
  where
  $\langle\ast,\ast\rangle:\widetilde{Q}\times\widetilde{Q}\to\mathbb{R}$
  stands for the real inner-product defined by
  \begin{align}
    \langle f,g\rangle=\int_{\mathbb{R}^{d}}dx\,f^{a}(x)g^{a}(x)\label{eq:A.5}
  \end{align}
  for arbitrary
  $f=(f^{1},\cdots,f^{n}),g=(g^{1},\cdots,g^{n})\in\widetilde{Q}$.
  The first term $(1/2)\langle\phi,C_{\Lambda}^{-1}\phi\rangle$ of
  \eqref{eq:A.4} is the kinetic term of the theory, where the operator
  $C_{\Lambda}$ and its inverse $C_{\Lambda}^{-1}$ stand for the
  covariance (i.e., cutoff propagator) and minus the cutoff Laplacian,
  respectively. More precisely, the operators $C_{\Lambda}$ and
  $C_{\Lambda}^{-1}$ are defined through the following integral
  transforms:
  \begin{subequations}
    \begin{align}
      (C_{\Lambda}\phi)^{a}(x)&=\int_{\mathbb{R}^{d}}dy\,C_{\Lambda}^{ab}(x,y)\phi^{b}(y),\label{eq:A.6a}\\
      (C_{\Lambda}^{-1}\phi)^{a}(x)&=\int_{\mathbb{R}^{d}}dy\,C_{\Lambda}^{-1\,ab}(x,y)\phi^{b}(y),\label{eq:A.6b}
    \end{align}
  \end{subequations}
  where $C_{\Lambda}^{ab}(x,y)$ and $C_{\Lambda}^{-1\,ab}(x,y)$ are
  $n\times n$-matrix-valued integral kernels defined by
  \begin{subequations}
    \begin{align}
      C_{\Lambda}^{ab}(x,y)&=\delta^{ab}\int_{\mathbb{R}^{d}}\frac{dp}{(2\pi)^{d}}\frac{K(\frac{|p|}{\Lambda})}{p^{2}}\e^{ip(x-y)},\label{eq:A.7a}\\
      C_{\Lambda}^{-1\,ab}(x,y)&=\delta^{ab}\int_{\mathbb{R}^{d}}\frac{dp}{(2\pi)^{d}}\frac{p^{2}}{K(\frac{|p|}{\Lambda})}\e^{ip(x-y)}.\label{eq:A.7b}
    \end{align}
  \end{subequations}
  Here $K(s)$ is a monotone decreasing cutoff function satisfying
  $K(0)=1$. We assume that $K(s)$ remains close to $1$ for
  $s\lesssim1$ and decays rapidly to $0$ for $s\gtrsim1$. Finally, the
  second term $V_{\Lambda,\Lambda_{0}}[\phi]$ of \eqref{eq:A.4} is a
  functional of $\phi$ and stands for an interaction term (including
  the mass term) of the theory.
  
  Now, the canonical partition function with respect to the Wilson
  action $S_{\Lambda^{\prime},\Lambda_{0}}[\phi]$ is given by the
  functional integral
  $Z=\int_{\widetilde{Q}}d\phi\e^{-S_{\Lambda^{\prime},\Lambda_{0}}[\phi]}$,
  where $\int_{\widetilde{Q}}d\phi$ is formally defined by
  \begin{align}
    \int_{\widetilde{Q}}d\phi=\prod_{x\in\mathbb{R}^{d}}\int_{\mathbb{R}^{n}}d\phi(x).\label{eq:A.8}
  \end{align}
  Notice that the Boltzmann weight
  $\e^{-S_{\Lambda^{\prime},\Lambda_{0}}[\phi]}\propto\e^{-\frac{1}{2}\langle\phi,C_{\Lambda^{\prime}}^{-1}\phi\rangle}=\exp(-\frac{1}{2}\int_{\mathbb{R}^{d}}\frac{dp}{(2\pi)^{d}}\frac{p^{2}|\Tilde{\phi}(p)|^{2}}{K(|p|/\Lambda^{\prime})})$
  exponentially tends to zero for $|p|/\Lambda^{\prime}\gtrsim1$ so
  that the functional integral
  $\int_{\widetilde{Q}}d\phi\e^{-S_{\Lambda^{\prime},\Lambda_{0}}[\phi]}$
  effectively reduces to the functional integral only over the
  momentum modes $\{\Tilde{\phi}(p)\}_{|p|\lesssim\Lambda^{\prime}}$,
  where $\Tilde{\phi}(p)=\int_{\mathbb{R}^{d}}dx\,\phi(x)\e^{-ipx}$ is
  the Fourier transform of $\phi(x)$. The
  partition-function-preserving ERG transformation
  $\mathsf{R}_{\Lambda,\Lambda^{\prime}}$ is easily constructed by
  integrating out the high-momentum modes
  $\{\Tilde{\phi}(p)\}_{\Lambda\lesssim|p|\lesssim\Lambda^{\prime}}$
  in the functional integral
  $\int_{\widetilde{Q}}d\phi\e^{-S_{\Lambda^{\prime},\Lambda_{0}}[\phi]}$. This
  can be done by using the so-called \textit{splitting formula}:
  \begin{align}
    \int_{\widetilde{Q}}d\phi\,\frac{\e^{-\frac{1}{2}\langle\phi,A^{-1}\phi\rangle}}{\sqrt{\det(2\pi A)}}F[\phi]=\int_{\widetilde{Q}}d\phi_{1}\frac{\e^{-\frac{1}{2}\langle\phi_{1},A_{1}^{-1}\phi_{1}\rangle}}{\sqrt{\det(2\pi A_{1})}}\int_{\widetilde{Q}}d\phi_{2}\frac{\e^{-\frac{1}{2}\langle\phi_{2},A_{2}^{-1}\phi_{2}\rangle}}{\sqrt{\det(2\pi A_{2})}}F[\phi_{1}+\phi_{2}],\label{eq:A.9}
  \end{align}
  where $F$ is an arbitrary functional of $\phi$. $A$, $A_{1}$ and
  $A_{2}$ are arbitrary positive-definite symmetric operators
  satisfying $A=A_{1}+A_{2}$. (For the proof of \eqref{eq:A.9}, see,
  e.g., Section 1.2.3 of \cite{Salmhofer:1999uq}. See also Section
  16.1 of \cite{Zinn-Justin:2007uvz} and \S2.1.1 of
  \cite{Igarashi:2009tj}.) The canonical partition function with
  respect to the Wilson action
  $S_{\Lambda^{\prime},\Lambda_{0}}[\phi]$ is then rewritten as
  \begin{align}
    Z
    &=\int_{\widetilde{Q}}d\phi\e^{-S_{\Lambda^{\prime},\Lambda_{0}}[\phi]}\nonumber\\
    &=\sqrt{\det(2\pi C_{\Lambda^{\prime}})}\int_{\widetilde{Q}}d\phi\,\frac{\e^{-\frac{1}{2}\langle\phi,C_{\Lambda^{\prime}}^{-1}\phi\rangle}}{\sqrt{\det(2\pi C_{\Lambda^{\prime}})}}\e^{-V_{\Lambda^{\prime},\Lambda_{0}}[\phi]}\nonumber\\
    &=\sqrt{\det(2\pi C_{\Lambda^{\prime}})}
      \int_{\widetilde{Q}}d\phi_{1}\,\frac{\e^{-\frac{1}{2}\langle\phi_{1},C_{\Lambda}^{-1}\phi_{1}\rangle}}{\sqrt{\det(2\pi C_{\Lambda})}}
      \int_{\widetilde{Q}}d\phi_{2}\,\frac{\e^{-\frac{1}{2}\langle\phi_{2},(C_{\Lambda^{\prime}}-C_{\Lambda})^{-1}\phi_{2}\rangle}}{\sqrt{\det(2\pi (C_{\Lambda^{\prime}}-C_{\Lambda}))}}
      \e^{-V_{\Lambda^{\prime},\Lambda_{0}}[\phi_{1}+\phi_{2}]}\nonumber\\
    &=\int_{\widetilde{Q}}d\phi_{1}\e^{-S_{\Lambda,\Lambda_{0}}[\phi_{1}]},\label{eq:A.10}
  \end{align}
  where
  \begin{align}
    \e^{-S_{\Lambda,\Lambda_{0}}[\phi]}=\int_{\widetilde{Q}}d\phi_{2}\frac{\e^{-\frac{1}{2}\langle\phi,C_{\Lambda}^{-1}\phi\rangle-\frac{1}{2}\langle\phi_{2},(C_{\Lambda^{\prime}}-C_{\Lambda})^{-1}\phi_{2}\rangle}}{\sqrt{\det(2\pi C_{\Lambda}C_{\Lambda^{\prime}}^{-1}(C_{\Lambda^{\prime}}-C_{\Lambda}))}}\e^{-V_{\Lambda^{\prime},\Lambda_{0}}[\phi+\phi_{2}]}.\label{eq:A.11}
  \end{align}
  Notice that the factor
  $\e^{-\frac{1}{2}\langle\phi_{2},(C_{\Lambda^{\prime}}-C_{\Lambda})^{-1}\phi_{2}\rangle}=\exp(-\frac{1}{2}\int_{\mathbb{R}^{d}}\frac{dp}{(2\pi)^{d}}\frac{p^{2}|\Tilde{\phi}_{2}(p)|^{2}}{K(|p|/\Lambda^{\prime})-K(|p|/\Lambda)})$
  is close to zero for $|p|\lesssim\Lambda$ and exponentially tends to
  zero for $|p|\gtrsim\Lambda^{\prime}$ so that the
  $\phi_{2}$-integral effectively reduces to the integration only over
  the high-momentum modes
  $\{\Tilde{\phi}_{2}(p)\}_{\Lambda\lesssim|p|\lesssim\Lambda^{\prime}}$. The
  $\phi_{1}$-integral, on the other hand, corresponds to the
  integration over the low-momentum modes
  $\{\Tilde{\phi}_{1}(p)\}_{0<|p|\lesssim\Lambda}$.

  Now we are ready to find the ERG kernel. Let us rewrite the
  functional integral \eqref{eq:A.11} as follows:
  \begin{align}
    \e^{-S_{\Lambda,\Lambda_{0}}[\phi]}
    &=\int_{\widetilde{Q}}d\phi^{\prime}\,\frac{\e^{-\frac{1}{2}\langle\phi,C_{\Lambda}^{-1}\phi\rangle-\frac{1}{2}\langle\phi^{\prime}-\phi,(C_{\Lambda^{\prime}}-C_{\Lambda})^{-1}(\phi^{\prime}-\phi)\rangle}}{\sqrt{\det(2\pi C_{\Lambda}C_{\Lambda^{\prime}}^{-1}(C_{\Lambda^{\prime}}-C_{\Lambda}))}}\e^{-V_{\Lambda^{\prime},\Lambda_{0}}[\phi^{\prime}]}\nonumber\\
    &=\int_{\widetilde{Q}}d\phi^{\prime}\,\frac{\e^{-\frac{1}{2}\langle\phi,C_{\Lambda}^{-1}\phi\rangle-\frac{1}{2}\langle\phi^{\prime}-\phi,(C_{\Lambda^{\prime}}-C_{\Lambda})^{-1}(\phi^{\prime}-\phi)\rangle+\frac{1}{2}\langle\phi^{\prime},C_{\Lambda^{\prime}}^{-1}\phi^{\prime}\rangle}}{\sqrt{\det(2\pi C_{\Lambda}C_{\Lambda^{\prime}}^{-1}(C_{\Lambda^{\prime}}-C_{\Lambda}))}}
      \e^{-\frac{1}{2}\langle\phi^{\prime},C_{\Lambda^{\prime}}^{-1}\phi^{\prime}\rangle-V_{\Lambda^{\prime},\Lambda_{0}}[\phi^{\prime}]}\nonumber\\
    &=\int_{\widetilde{Q}}d\phi^{\prime}\,\widetilde{\mathsf{R}}_{\Lambda,\Lambda^{\prime}}[\phi,\phi^{\prime}]\e^{-S_{\Lambda^{\prime},\Lambda_{0}}[\phi^{\prime}]},\label{eq:A.12}
  \end{align}
  where
  \begin{align}
    \widetilde{\mathsf{R}}_{\Lambda,\Lambda^{\prime}}[\phi,\phi^{\prime}]
    &=\frac{\e^{-\frac{1}{2}\langle\phi,C_{\Lambda}^{-1}\phi\rangle-\frac{1}{2}\langle\phi^{\prime}-\phi,(C_{\Lambda^{\prime}}-C_{\Lambda})^{-1}(\phi^{\prime}-\phi)\rangle+\frac{1}{2}\langle\phi^{\prime},C_{\Lambda^{\prime}}^{-1}\phi^{\prime}\rangle}}{\sqrt{\det(2\pi C_{\Lambda}C_{\Lambda^{\prime}}^{-1}(C_{\Lambda^{\prime}}-C_{\Lambda}))}}\nonumber\\
    &=\frac{\e^{-\frac{1}{2}\langle\phi-C_{\Lambda}C_{\Lambda^{\prime}}^{-1}\phi^{\prime},C_{\Lambda}^{-1}C_{\Lambda^{\prime}}(C_{\Lambda^{\prime}}-C_{\Lambda})^{-1}(\phi-C_{\Lambda}C_{\Lambda^{\prime}}^{-1}\phi^{\prime})\rangle}}{\sqrt{\det(2\pi C_{\Lambda}C_{\Lambda^{\prime}}^{-1}(C_{\Lambda^{\prime}}-C_{\Lambda}))}}.\label{eq:A.13}
  \end{align}
  This is the ERG kernel on $\widetilde{Q}$. As easily checked, this
  kernel satisfies the following properties for any
  $\phi,\phi^{\prime},\phi^{\prime\prime}\in\widetilde{Q}$,
  $\Lambda\leq\Lambda^{\prime}\leq\Lambda^{\prime\prime}(\leq\Lambda_{0})$,
  and $(a,R)\in\mathbb{R}^{n}\rtimes O(n)\cong\Isom(\mathbb{R}^{n})$:
  \begin{subequations}
    \begin{itemize}
    \item\textbf{Property 1 (Composition law).}
      \begin{align}
        \int_{\widetilde{Q}}d\phi^{\prime}\,\widetilde{\mathsf{R}}_{\Lambda,\Lambda^{\prime}}[\phi,\phi^{\prime}]\widetilde{\mathsf{R}}_{\Lambda^{\prime},\Lambda^{\prime\prime}}[\phi^{\prime},\phi^{\prime\prime}]=\widetilde{\mathsf{R}}_{\Lambda,\Lambda^{\prime\prime}}[\phi,\phi^{\prime\prime}].\label{eq:A.14a}
      \end{align}
    \item\textbf{Property 2 (Initial condition).}
      \begin{align}
        \widetilde{\mathsf{R}}_{\Lambda,\Lambda}[\phi,\phi^{\prime}]=\delta[\phi-\phi^{\prime}].\label{eq:A.14b}
      \end{align}
    \item\textbf{Property 3 ($\Isom(\mathbb{R}^{n})$-invariance).}
      \begin{align}
        \widetilde{\mathsf{R}}_{\Lambda,\Lambda^{\prime}}[(a,R)\phi,(a,R)\phi^{\prime}]=\widetilde{\mathsf{R}}_{\Lambda,\Lambda^{\prime}}[\phi,\phi^{\prime}].\label{eq:A.14c}
      \end{align}
    \item\textbf{Property 4 (Flow equation).}
      \begin{align}
        -\Lambda\frac{\partial}{\partial\Lambda}\widetilde{\mathsf{R}}_{\Lambda,\Lambda^{\prime}}[\phi,\phi^{\prime}]=\mathsf{G}_{\Lambda}[\phi,\tfrac{\delta}{\delta\phi}]\widetilde{\mathsf{R}}_{\Lambda,\Lambda^{\prime}}[\phi,\phi^{\prime}].\label{eq:A.14d}
      \end{align}
    \item\textbf{Property 5 (Reflection reality).}
      \begin{align}
        \overline{\widetilde{\mathsf{R}}_{\Lambda,\Lambda^{\prime}}[\widetilde{\vartheta}\phi,\widetilde{\vartheta}\phi^{\prime}]}=\widetilde{\mathsf{R}}_{\Lambda,\Lambda^{\prime}}[\phi,\phi^{\prime}].\label{eq:A.14e}
      \end{align}
    \item\textbf{Property 6 (Normalization condition).}
      \begin{align}
        \int_{\widetilde{Q}}d\phi\,\widetilde{\mathsf{R}}_{\Lambda,\Lambda^{\prime}}[\phi,\phi^{\prime}]=1.\label{eq:A.14f}
      \end{align}
    \end{itemize}
  \end{subequations}
  Here $(a,R)\phi$ in \eqref{eq:A.14c} stands for the action of
  $\Isom(\mathbb{R}^{n})$ on $\phi\in\widetilde{Q}$ defined by
  \begin{align}
    ((a,R)\phi)^{a}(x)=a^{a}+R^{ab}\phi^{b}(x),\label{eq:A.16}
  \end{align}
  where $a=(a^{1},\cdots,a^{n})\in\mathbb{R}^{n}$ and
  $R=(R^{ab})\in O(n)$ are an arbitrary $n$-vector and an orthogonal
  matrix. In \eqref{eq:A.14d},
  $\mathsf{G}_{\Lambda}[\phi,\delta/\delta\phi]$ is the
  functional-differential-operator realization of the ERG generator
  given by
  \begin{align}
    \mathsf{G}_{\Lambda}[\phi,\tfrac{\delta}{\delta\phi}]=-\frac{1}{2}\left\langle\frac{\delta}{\delta\phi},\Dot{C}_{\Lambda}\frac{\delta}{\delta\phi}\right\rangle-\left\langle\frac{\delta}{\delta\phi},C_{\Lambda}^{-1}\Dot{C}_{\Lambda}\phi\right\rangle,
    \quad
    \Dot{C}_{\Lambda}=-\Lambda\frac{\partial}{\partial\Lambda}C_{\Lambda},\label{eq:A.15}
  \end{align}
  where for any (symmetric) operator $A$ whose integral kernel is
  $A^{ab}(x,y)$, the inner products
  $\langle\frac{\delta}{\delta\phi},A\frac{\delta}{\delta\phi}\rangle$
  and $\langle\frac{\delta}{\delta\phi},A\phi\rangle$ stand for the
  functional-differential operators defined by
  \begin{subequations}
    \begin{align}
      \left\langle\frac{\delta}{\delta\phi},A\frac{\delta}{\delta\phi}\right\rangle
      &=\int_{\mathbb{R}^{d}}dx\int_{\mathbb{R}^{d}}dy\,\frac{\delta}{\delta\phi^{a}(x)}A^{ab}(x,y)\frac{\delta}{\delta\phi^{b}(y)},\label{eq:A.17a}\\
      \left\langle\frac{\delta}{\delta\phi},A\phi\right\rangle
      &=\int_{\mathbb{R}^{d}}dx\int_{\mathbb{R}^{d}}dy\,\frac{\delta}{\delta\phi^{a}(x)}A^{ab}(x,y)\phi^{b}(y).\label{eq:A.17b}
    \end{align}
  \end{subequations}
  Note that
  $\langle\frac{\delta}{\delta\phi},A\phi\rangle=\langle
  A\phi,\frac{\delta}{\delta\phi}\rangle+\tr(A)$, where
  $\langle
  A\phi,\frac{\delta}{\delta\phi}\rangle=\int_{\mathbb{R}^{d}}\!dx\int_{\mathbb{R}^{d}}\!dy\,A^{ab}(x,y)\phi^{b}(y)\frac{\delta}{\delta\phi^{a}(x)}$
  and
  $\tr(A)=\sum_{a=1}^{n}\int_{\mathbb{R}^{d}}\!dx\,A^{aa}(x,x)$. Finally,
  $\widetilde{\vartheta}\phi$ in \eqref{eq:A.14e} is defined by
  $(\widetilde{\vartheta}\phi)(x)=\phi(rx)$, where $rx$ stands for the
  reflection of $x=(x^{1},\cdots,x^{d})$ given, e.g., by
  $rx=(-x^{1},x^{2},\cdots,x^{d})$.

  \subsection{ERG kernel on
    \texorpdfstring{$Q=\Map(\mathbb{R}^{d},\mathbb{R}^{n}/\Gamma)$}{Map(Rd,Rn/Gamma)}}
  \label{appendix:A.2}
  Now it is easy to construct the ERG kernel for the $d$-dimensional
  Euclidean scalar field theory whose target space is the compact flat
  Euclidean manifold $\mathbb{R}^{n}/\Gamma$. The result is
  \begin{align}
    \mathsf{R}_{\Lambda,\Lambda^{\prime}}[\phi,\phi^{\prime}]=\sum_{\gamma\in\Gamma}\widetilde{\mathsf{R}}_{\Lambda,\Lambda^{\prime}}[\phi,\gamma\phi^{\prime}].\label{eq:A.18}
  \end{align}
  One can easily show that this kernel satisfies the composition law,
  the initial condition, the reflection reality, and the normalization
  condition. Note that the kernel \eqref{eq:A.18} satisfies the same
  flow equation as \eqref{eq:A.14d}.
\end{appendices}

\printbibliography[heading=bibintoc]
\end{document}